\documentclass[journal]{IEEEtran}

\usepackage{amsmath}
\allowdisplaybreaks[4]
\usepackage{amssymb}
\usepackage{amsthm}
\usepackage{cases}
\usepackage{bm}
\usepackage{algorithm}
\usepackage{algorithmic}
\usepackage{cases}
\usepackage{subfigure}
\usepackage{multirow}
\usepackage{xcolor} 
\usepackage{enumerate}
\usepackage{etoolbox}
\usepackage{boxedminipage} 
\usepackage{booktabs} 
\usepackage{graphicx}
\usepackage{mathrsfs}
\usepackage{caption}
\usepackage[pagebackref=true,breaklinks=true,letterpaper=true,colorlinks,bookmarks=false]{hyperref}

\newcommand{\orcid}[1]{\href{https://orcid.org/#1}{\includegraphics[scale=0.5]{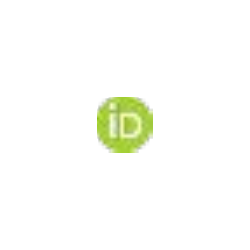}}}

\newcommand{\Chunyan}[1]{{\color{black}#1}} 
\newcommand{\Jiatao}[1]{{\color{black}#1}} 
\newcommand{\Jiang}[1]{{\color{black}#1}} 
\newcommand{\JJ}[1]{{\color{black}#1}} 

\newcommand{\MyMapTemplatePrefix}[4]{\expandafter#1\csname#3#4\endcsname{#2{#4}}}
\newcommand{\MyMapTemplatePrefixNew}[5]{\expandafter#1\csname#4#5\endcsname{#2{#3{#5}}}}
\forcsvlist{\MyMapTemplatePrefix {\def} {\mathbf} {}} {A,B,C,D,E,F,G,H,I,J,K,L,M,N,O,P,Q,R,S,T,U,V,W,X,Y,Z} 
\forcsvlist{\MyMapTemplatePrefix {\def} {\mathbf} {}} {a,b,c,d,e,f,g,h,j,k,l,m,n,o,p,q,r,s,t,u,v,w,x,y,z,1,0}
\forcsvlist{\MyMapTemplatePrefix {\def} {\widetilde} {wt}} {A,B,C,D,E,F,G,H,I,J,K,L,M,N,O,P,Q,R,S,T,U,V,W,X,Y,Z}
\forcsvlist{\MyMapTemplatePrefix {\def} {\widetilde} {wt}} {a,b,c,d,e,f,g,h,i,j,k,l,m,n,o,p,q,r,s,t,u,v,w,x,y,z} 
\forcsvlist{\MyMapTemplatePrefixNew {\def} {\widetilde}{\mathbf} {tb}} {A,B,C,D,E,F,G,H,I,J,K,L,M,N,O,P,Q,R,S,T,U,V,W,X,Y,Z}
\forcsvlist{\MyMapTemplatePrefixNew {\def} {\widetilde}{\mathbf} {tb}} {a,b,c,d,e,f,g,h,i,j,k,l,m,n,o,p,q,r,s,t,u,v,w,x,y,z}
\forcsvlist{\MyMapTemplatePrefix {\def} {\widehat} {wh}} {A,B,C,D,E,F,G,H,I,J,K,L,M,N,O,P,Q,R,S,T,U,V,W,X,Y,Z}
\forcsvlist{\MyMapTemplatePrefix {\def} {\widehat} {wh}} {a,b,c,d,e,f,g,h,i,j,k,l,m,n,o,p,q,r,s,t,u,v,w,x,y,z}
\forcsvlist{\MyMapTemplatePrefixNew {\def} {\widehat}{\mathbf} {hb}} {A,B,C,D,E,F,G,H,I,J,K,L,M,N,O,P,Q,R,S,T,U,V,W,X,Y,Z}
\forcsvlist{\MyMapTemplatePrefixNew {\def} {\widehat}{\mathbf} {hb}} {a,b,c,d,e,f,g,h,i,j,k,l,m,n,o,p,q,r,s,t,u,v,w,x,y,z}
\forcsvlist{\MyMapTemplatePrefix {\def} {\mathcal}{mc}} {A,B,C,D,E,F,G,H,I,J,K,L,M,N,O,P,Q,R,S,T,U,V,W,X,Y,Z,a,b,c,d,e,f,g,h,i,j,k,l,m,n,o,p,q,r,s,t,u,v,w,x,y,z}
\forcsvlist{\MyMapTemplatePrefix {\def} {\mathbb} {mb}} {A,B,C,D,E,F,G,H,I,J,K,L,M,N,O,P,Q,R,S,T,U,V,W,X,Y,Z}
\forcsvlist{\MyMapTemplatePrefix {\DeclareMathOperator} {} {} } {tr,diag,sgn}
 \def\st{\text{s.t.~}}
\def\ie{{i.e.}} \def\etal{{et. al}}
 \def\eg{{e.g.}}
\def\bSigma{\mathbf{\Sigma}} \def\bmu{\mathbf{\mu}}   \def\bTheta{\mathbf{\Theta}}
\def\bsigma{\mathbf{\sigma}}
\def\balpha{\mathbf{\alpha}}

\def\balpha{\boldsymbol{\alpha}}

\begin{document}
%
\title{Walk-Steered Convolution for Graph Classification}
%
%
%


\author{
Jiatao~Jiang\orcid{0000-0001-8656-877X}, 
Chunyan~Xu\orcid{0000-0002-0814-4362}, 
Zhen~Cui\orcid{0000-0002-8292-6389}, \IEEEmembership{Member, IEEE}, 
Tong~Zhang\orcid{0000-0001-6212-4891}, \IEEEmembership{Member, IEEE}, \\
Wenming~Zheng\orcid{0000-0002-7764-5179}, \IEEEmembership{Senior Member, IEEE}, 
and Jian~Yang\orcid{0000-0003-4800-832X}, \IEEEmembership{Senior Member, IEEE}

\thanks{ 
This work was supported by the National Natural Science Foundation of China under Grant 61772276, Grant 61972204, Grant 61906094, Grant 61921004 and Grant U1713208,
in part by the Natural Science Foundation of Jiangsu Province under Grant BK20190019 and Grant BK20191283,
in part by the Jiangsu Provincial Key Research and Development Program under Grant BE2016768,
and in part by the National Key Research and Development Program of China under Grant 2018YFB1305200.
\textit{(Jiatao Jiang and Chunyan Xu contributed equally to this work.)(Corresponding author: Zhen Cui.)}}

\thanks{J. Jiang, C. Xu, Z. Cui, T. Zhang and J. Yang are the Key Laboratory of Intelligent Perception and Systems for High-Dimensional Information of the Ministry of Education, School of Computer Science and Engineering, Nanjing University of Science and Technology, Nanjing 210094, China, and also with the Jiangsu Key Lab of Image and Video Understanding for Social Security, School of Computer Science and Engineering, Nanjing University of Science and Technology, Nanjing 210094, China (e-mail: jiatao@njust.edu.cn; cyx@njust.edu.cn; zhen.cui@njust.edu.cn; tong.zhang@njust.edu.cn; csjyang@njust.edu.cn).}

\thanks{W. Zheng is with the Key Laboratory of Child Development and Learning Science of the Ministry of Education, School of Biological Science and Medical Engineering, Southeast University, Nanjing 210096, China	(e-mail: wenming\_zheng@seu.edu.cn).}


}

%
%



\markboth{IEEE TRANSACTIONS ON NEURAL NETWORKS AND LEARNING SYSTEMS, November~2019}%
{Shell \MakeLowercase{\textit{et al.}}: Bare Demo of IEEEtran.cls for IEEE Journals}

%



\maketitle

\begin{abstract}
Graph classification is a fundamental but challenging issue for numerous real-world applications. Despite recent great progress in image/video classification, \Jiang{convolutional neural networks (CNNs)} cannot yet cater to graphs well because of graphical non-Euclidean topology. 
In this work, we propose a walk-steered convolutional (WSC) network to \JJ{assemble} the essential success of standard convolutional neural networks as well as the powerful representation ability of random walk. 
Instead of deterministic neighbor searching used in previous graphical CNNs, we construct multi-scale walk fields (a.k.a. local receptive fields) with random walk paths to depict subgraph structures and advocate graph scalability. 
To express the internal variations of a walk field, \Jiang{Gaussian mixture models} are introduced to encode principal components of walk paths therein. 
As an analogy to a standard convolution kernel on image, Gaussian models implicitly coordinate those unordered \Jiang{vertices/nodes and edges} in a local receptive field after projecting \Jiang{to the gradient space of Gaussian parameters.}
We further stack graph coarsening upon Gaussian encoding by using \Jiang{dynamic} clustering, such that high-level semantics of graph can be well learned like the conventional pooling on image. The experimental results on several public datasets demonstrate the superiority of our proposed WSC method over many state-of-the-arts for graph classification.
\end{abstract}


%
\IEEEpeerreviewmaketitle

\section{Introduction}
\IEEEPARstart{G}{raph} as a popular representation of irregular/non-Euclidean data is \Jiang{widely used} to model and analyze irregular/non-Euclidean data, such as social network, biological protein-protein interaction network, molecular graph structures, text data and so on. 
The graph-structured sample consists of a finite set of vertices/nodes, together with a set of connections revealing the relationship between unordered
pairs of these vertices (named edges).
A crucial problem therein is to annotate labels for graphs, which we broadly name as graph classification. For examples, for protein or \JJ{enzyme} data, we might be interested in discovering the \JJ{apparitions} of diseases or defective compounds, or for in a social network, we might be interested in predicting the interests of users.

\Chunyan{Driven by deep learning~\cite{NIPS2012_imagenet,googleLeNet,resnet}}, recent \JJ{convolutional} neural networks (CNNs) on graphs \Chunyan{have raised} a promising direction to graph classification~\cite{goyal2017graph,hamilton2017representation,jiang2018gaussian}. 
Generally, they fall into two categories: spectral methods~\cite{henaff2015deep} and spatial methods~\cite{orsini2018shift,niepert2016learning}. 
The former attempts to use or develop new spectral graph techniques for graphs by \JJ{virtues} of the classic signal theory. 
However, the computation of eigenvalue decomposition often burdens these spectral methods. 
Accordingly, local approximation is introduced into some recent variants~\cite{defferrard2016convolutional,atwood2016diffusion} in order to accelerate the \Jiang{filtering}. 
For the latter, local \JJ{neighbor} relationships are usually aggregated spatially via edge connections for the sake of consistent responses on homomorphic graph structures. However, the plain aggregation or diffusion on edges will result into tied filtering on \Jiang{local neighbors}, thus in principle local variations of a receptive field region can not be well-encoded in all these convolution methods~\cite{orsini2018shift,defferrard2016convolutional,atwood2016diffusion}. 
In contrast, a $3 \times 3$ \JJ{convolution} kernel on image actually employs 9 different mappings on different spatial positions (e.g., top-left, top, top-right, etc.). 
Just due to the position-associated filtering, local appearance variations of an image are well-expressed in the conventional CNNs. 
Analogous to this, the recent PSCN~\cite{niepert2016learning} attempted to strictly sort and prune \Jiang{neighbors}, and then \Chunyan{defined} untied filters like a standard \JJ{convolution} kernel. 
But for flexible \Jiang{structures}, such an overly strict way in PSCN is more prone to be affected by \Jiang{neighbor} sizes and graph noises. 
Furthermore, as the deterministic way to local neighbors, these convolutional methods usually lack a good scalability to graphs with different sizes.

To address these above problems, in this paper, we propose a random walk based convolutional \Jiang{neural} network, named walk-steered \Jiang{convolutional (WSC) network}, for the general graph classification task. 
\Chunyan{Random walk~\cite{ng2016fast,wu2014cotransfer} has been employed to depict} various properties of graph because of its flexible and stochastic manner. 
Especially, graphical topology structures can be well preserved according to the theory of random walk. 
Motivated by this point, a local receptive field (or centralized subgraph) around one vertex can be defined as a walk field composed of \Jiang{walk paths (or walks)} with a fixed length starting at this vertex. 
Thus \Jiang{a local receptive field} is converted to the set of walk paths therein. 
The variations of local receptive field are embodied in the distribution of walk paths, each of which is associated with attributes of its vertices besides the path information. 
By varying the walk lengths, we can further construct multi-scale walk fields to better express the local \JJ{neighbor} relationship of graph data.

\begin{figure*}[!t]
	\centering
	\includegraphics[width=0.9\textwidth]{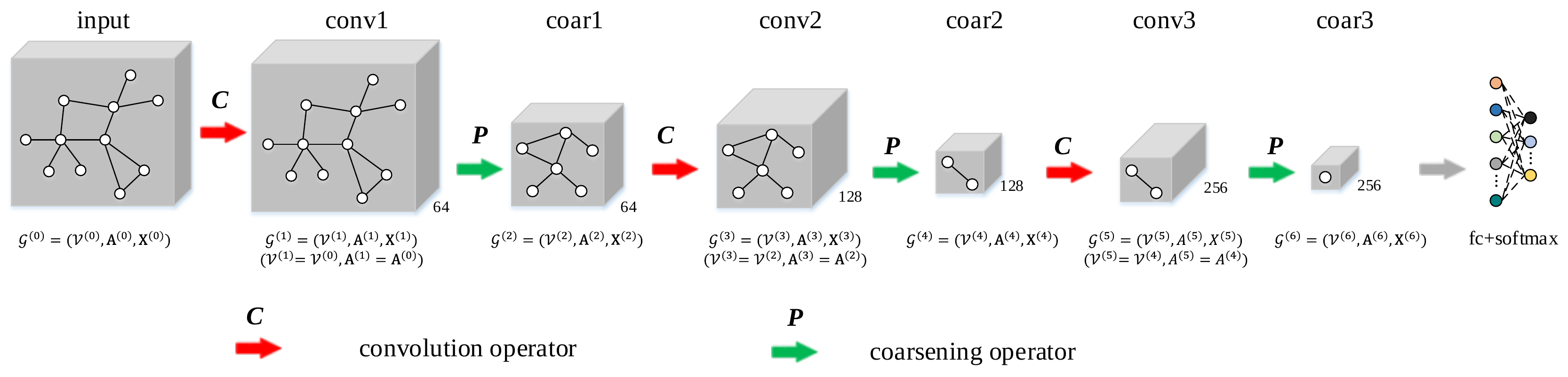}
	\caption{\Chunyan{An illustration of our proposed WSC network. 
	Given an input graph data, WSC can predict the sample-wise category in an end-to-end way. Like standard CNNs, the WSC network mainly consists of stacked \Jiang{graph convolution layers, coarsening layers and a fully connected (FC) layer with a final softmax layer.} 
    The \JJ{detailed} network configuration can be found in Table~\ref{table:config}. The graph convolution operator can aggregate \JJ{neighbor} information \JJ{to generate new convolutional features within the local receptive field for all vertices.} The graph convolution operator can still keep the structure of original graph, as described in Section~\ref{sec:rf} and \ref{sec:gaussian}. The graph coarsening operator clusters some local vertices to generate a hierarchical graph, which is presented in Section~\ref{sec:coasen}.}}
    \label{fig:network}
	\vskip -0.1in
\end{figure*}
Encoding local variations is critical to convolution filtering just like standard CNNs. 
To this end, we introduce \Jiang{Gaussian mixture models} to characterize each centralized subgraph, \ie, the distribution of walk field. Thus those unordered walk paths will be coordinated to align the implicit ``directions" guided by Gaussian models.
On different ``directions", we can learn various convolutional filters, which should be the same to the position-associated convolution kernel on image, \eg, untied filter vectors respectively on nine spatial positions for a $3\times 3$ \JJ{convolution} kernel. 
For the solution of Gaussian models, we derive the gradients of Gaussian parameters as the output responses. 
The advantage of gradient calculation includes two folds: i) make the network infer forward easily; ii) avoid the complicated EM algorithm. A graph convolution layer is composed of the construction of walk fields and the Gaussian mixture encoding. 
We can further stack it as well as graph coarsening layer (\eg, using graph cut~\cite{dhillon2007weighted} or clustering algorithm~\cite{defferrard2016convolutional}) to a multi-layer graphical convolutional network. 
Therefore, our proposed graph convolution simultaneously \JJ{assembles} the flexibility and scalability of random walks to graphs, and the representation ability of Gaussian models to variations of unordered examples.

In summary, we propose a novel walk-steered graph convolution for graph classification. Concretely, the major contributions of this work can be summarized as follows:
\begin{itemize}
\item We introduce random \Jiang{walk} into the construction of local receptive fields on \JJ{graph-structured} data, named walk fields, which differ from previous deterministic convolution strategies with spectral or spatial methods.
\item Gaussian mixture models are adopted to encode local variations of receptive field, which can be different to those previous methods aggregating/sorting neighbor vertices.
\item Extensive experiments of graph classification demonstrate that our proposed WSC can achieve the state-of-the-art performance on most \JJ{social and bioinformatic network} datasets.
\end{itemize}

\section{Related Work}\label{sec:related-work}

\Chunyan{Convolutional neural networks have been successfully used in many science and business domains (including image recognition~\cite{googleLeNet,resnet}, video recognition~\cite{tsn} and autonomous driving~\cite{driving1}), when compared with these traditional machine learning methods~\cite{gou2018two,gou2019classifier}.} 
\Chunyan{A classic CNN (\eg, AlexNet~\cite{NIPS2012_imagenet}) is composed of alternatively stacked \JJ{convolution} layers and spatial pooling layers, followed with fully connected layers. 
The \JJ{convolution} layer is to extract feature maps by linear convolutional filters followed by nonlinear activation functions (e.g., rectifier, sigmoid, tanh, etc.). 
Spatial pooling layer is to group the local features together from spatially adjacent pixels, and is typically used to improve the robustness to the slight deformation of objects.} 
Recently, convolutional neural networks based methods \Chunyan{have been} paid more attention \Jiang{to} analyze the graph-structured data. 

\Chunyan{Graph CNN is a \Jiang{type of powerful} neural network designed to work directly on the graph-structured data, where various graph convolution and pooling operations have been studied to perform deep learning and then learn the discriminative representation on graph.}
For example, various graph convolution approaches~\cite{kashima2003marginalized,morris2017glocalized,shervashidze2009efficient,yanardag2015deep} have been proposed for dealing with graph-structured data over the past few decades. 
They can be divided into two main categories: spectral methods~\cite{henaff2015deep,kipf2016semi} and spatial methods~\cite{niepert2016learning,li2015gated}. 
Spectral methods~\cite{henaff2015deep} have constructed a series of spectral filters by decomposing graph Laplacian \JJ{matrix}, which can often lead to a high-computational burden on eigenvalue decomposition. 
To reduce computational complexity, more recently, the fast local spectral filtering method~\cite{defferrard2016convolutional} parameterized the frequency responses as a Chebyshev polynomial approximation.
However, after summarizing all nodes, this method will discard ``directions" of local receptive field.
For this, Such \etal~\cite{such2017robust} attempted to design multiple \Jiang{adjacency} matrices for each graph, which was actually non-trivial for general graphs. 
Furthermore, this kind of methods~\cite{kipf2016semi} usually required equal sizes of graphs like the same sizes of images for CNNs.
On the other side, spatial methods have attempted to explicitly model spatial structures of adjacent vertices through aggregating or sorting \Jiang{neighbors}. 
For examples, diffusion CNNs~\cite{atwood2016diffusion} scanned a diffusion process across each \JJ{vertex}; PSCN~\cite{niepert2016learning} linearized neighbors by sorting edges and then derived convolution on these sorted vertices; NgramCNN~\cite{luo2017deep} serialized each graph by introducing the concept of $n$-gram block; GraphSAGE~\cite{hamilton2017inductive} and EP-B~\cite{garcia2017learning} used the aggregation or propagation of local neighbor nodes. 
For the \Jiang{linearized local neighbors}, RNNs~\cite{li2015gated} was often employed to model the structured sequences.

Graph classification has gained increasing attention recently and various approaches~\cite{scarselli2009graph,wang2016graph,wang2017incremental,zhao2019hashing,liu2019si} have been developed to deal with the problem. 
\Chunyan{One classic technique line of graph classification is to use \Jiang{the local information on graphs}, where the construction process can be achieved by random walk~\cite{kashima2003marginalized,dong2017metapath2vec}, subtree~\cite{shervashidze2011weisfeiler}, graphlets~\cite{shervashidze2009efficient,prvzulj2007biological}, etc.
For example,
Zhao et al.~\cite{zhao2019hashing} have proposed to construct a regular subspace by a shared hash function to project the neighbors.
Gou et al.~\cite{gou2018sparsity} have developed a novel discriminative dimensionality reduction technique, named sparsity and
geometry preserving graph embedding, by constructing global and local adjacent graphs.
The metapath2vec model~\cite{dong2017metapath2vec}, which was proposed to learn the scalable representation of heterogeneous networks, formalized meta-path based random walks to construct the heterogeneous \JJ{neighbors} of a node and then leveraged a heterogeneous skip-gram model to perform node embeddings.
Przulj et al.~\cite{prvzulj2007biological} have proposed to obtain the similarity between biological networks by using graphlet degree distribution, which can employ
the systematic measure of a network's local structure that imposes a large number of similarity constraints on networks being compared.} 
However, the kernel methods often suffer a high complexity especially when facing the large-scale data. Recently the idea of deep learning has been introduced into graph kernel in the literature~\cite{yanardag2015deep}. More recently, the aforementioned spatial and spectral methods~\cite{henaff2015deep,niepert2016learning,defferrard2016convolutional,such2017robust,bruna2014spectral} can be also adopted for the graph classification task.
In contrast, our WSC is different from these existing approaches in two main aspects.
First, we use random walks to describe \Jiang{the} receptive field, which is more flexible to large-scale graphs or partially observed graphs.
Besides, \Chunyan{various random walk based methods~\cite{perozzi2014deepwalk,grover2016node2vec,ng2012co}
have been proposed to learn a node embedding representation} by preserving edge connectivity.
Different from these approaches, we conduct (multi-layer) convolution on random walks to abstract graph representation like the standard CNNs.
Second, we implicitly define principal ``directions" of \Jiang{the} receptive field by using \Jiang{Gaussian mixture models}, which encode variations of graph structures more accurately.
Recently, the geometric deep learning method~\cite{monti2017geometric} also introduced mixture model CNNs~\cite{monti2017geometric}.
Different from this work, which only adopts \JJ{Gaussian models} to weight neighbors, our WSC can derive Gaussian components of subgraphs.
\Chunyan{Moreover, the nearest neighbors-based graph convolution methods~\cite{monti2017geometric} are deterministic by considering neighbor nodes with a certain field, while our proposed graph convolution with walk fields can randomly sample to gain a rich walk receptive field for better capturing the graph topology structures.}

\section{Walk-steered convolution network}

\begin{figure}[!t]
	\centering
	\includegraphics[width=0.35\textwidth]{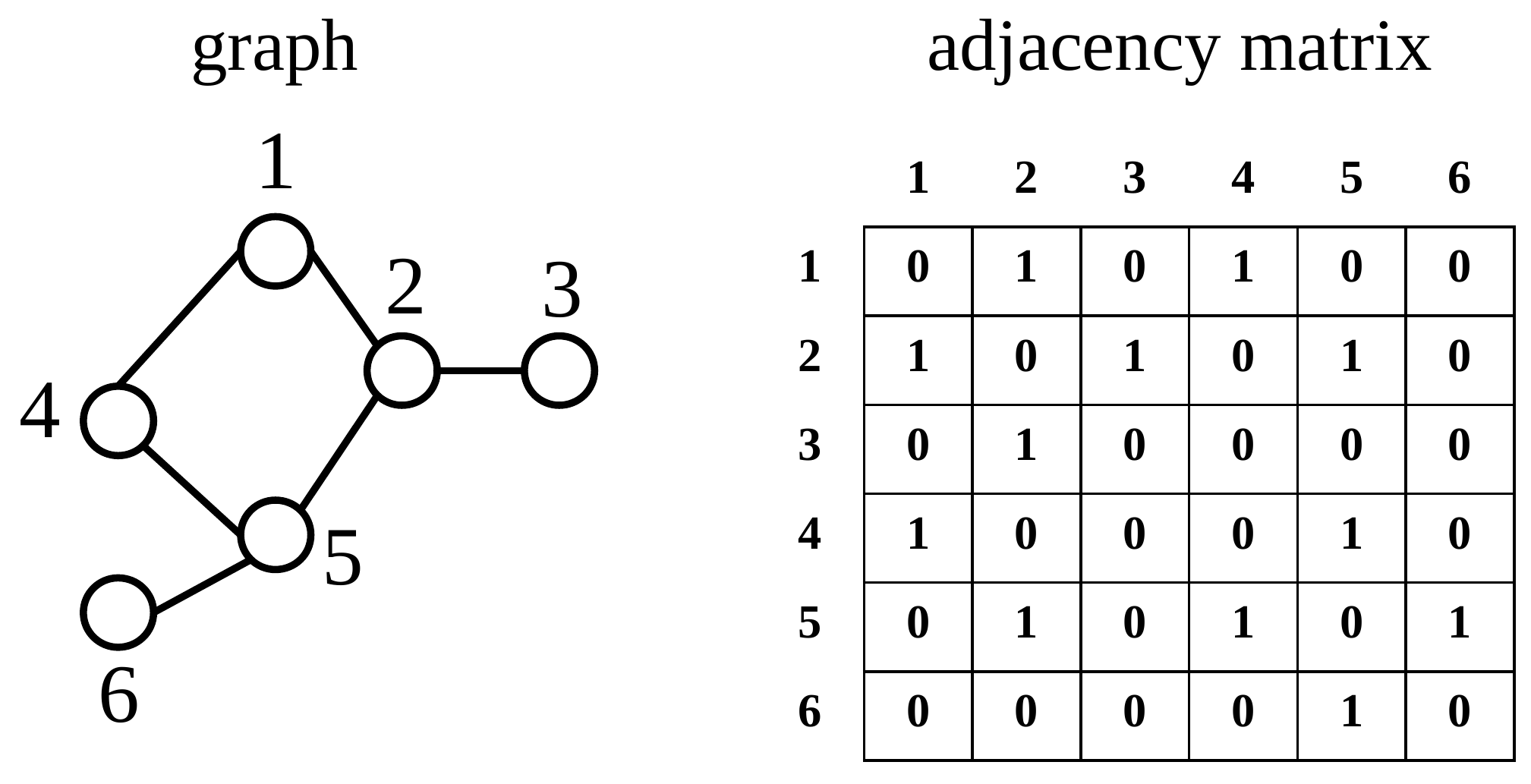}
	\caption{\Chunyan{An example of graph data and its corresponding adjacency matrix. The undirected graph consists of six vertices $v_1, \cdots, v_6$. The adjacency matrix of graph is a symmetric binary $6 \times 6$ matrix.}}	
	\label{fig:graph}
\end{figure}

We denote an undirected/directed graph with a triple $\mcG=(\mcV,\A,\X)$, where $\mcV = \{v_i\}_{i=1}^{m}$ is a finite set of $m$ vertices, $\A \in \mbR^{m \times m}$ is a (weighted) adjacency matrix and $\X \in \mbR^{m \times d}$ denotes an attribute (or signal) matrix of vertices. 
The adjacency matrix $\A$ records connection relationships between vertices. $A(v_i, v_j)\Jiang{\neq} 0$ if and only if $(v_i, v_j)\in \mcE$ (the set of edges), otherwise $A(v_i,v_j)\Jiang{=} 0$. For simplification, we also write $A(v_i, v_j)$ as $A_{ij}$. 
The attribute matrix $\X$ encloses the attributes of the vertices $\mcV$, where the $i$-th row $\X_{v_i}$ (or denoted $\X_{i\cdot}$) corresponds to the $d$-dimension attribute of vertex $v_i$. 
\Chunyan{Fig.~\ref{fig:graph} and Table~\ref{table:symbols} show the example of graph and the summary of symbols, respectively. }

\begin{table}[!h]
	\centering
	\footnotesize
	\caption{\Chunyan{Summary of symbols in this work}}
	\label{table:symbols}
	\begin{tabular}{c l}
		\toprule
		symbol  & description\\
		\midrule
		$\mathcal{G}$   &graph \\
		$\mathcal{V}$   &node/vertex \\
		$\mathbf{A}$    &adjacency matrix \\
		$\mathbf{X}$    &node attribute/feature \\
		$t/T$           &receptive field scale/hop  \\
		$P$             &transition probability between two vetices  \\
		$k/K$           &sample times of random walk \\   
		$c/C$           &number of Gaussian components  \\
		$\mathcal{P}$   &a walk path or a walk \\
		$\mathcal{W}$   &a walk field  \\
		$\bTheta=(w/\balpha, \mu, \mathbf{\Sigma})$        &parameter of Gaussian components  \\
		$\mathcal{N}$   &Gaussian function   \\
		$\mathcal{Q}$   &probability for Gaussian components \\
		$\zeta$         &log-likelihood function of the receptive field  \\
		$\mathscr{G}$   &gradient of Gaussian model parameters \\
		$\mathbf{F}$    &feature vector \\ 
		$\mathbf{P}$    &pooling matrix   \\
		$\gamma / \Gamma$  &vetex weight  \\
		$RW(\cdot; \cdot, \cdot)$            &random walk function \\
		$G(\cdot, \cdot)$             &Gaussian encoding function \\
		$f(\cdot), g(\cdot)$          &feature mapping function  \\
		$\psi(\cdot)$                   &clustering function   \\
		$\phi(\cdot)$                 &differentiable weight function  \\
		\bottomrule
	\end{tabular}
	\vskip -0.1in
\end{table}

\subsection{Network overview}

The graph convolution network consists of two basic layers: the convolution layer and the graph coarsening layer, similar to the standard CNNs. 
As shown in Fig.~\ref{fig:gaussianencoding}, the convolution layer is composed of two modules: the construction of walk fields (Section~\ref{sec:rf}) and the Gaussian mixture encoding (Section~\ref{sec:gaussian}). The coarsening layer takes a max-pooling-like strategy to select maximum responses from each group of clustered vertices. The purpose is to abstract high-level contours\Jiang{,} even semantics and simultaneously reduce the computational cost by massacring graphical vertices. The coarsening details can be found in Section~\ref{sec:coasen}. 
Thus we can alternatively stack the two layers into a multi-layer graph convolution network. With the increase of layers, the receptive field size of convolution filters will become larger, and the topper layer can extract more global information. 
In the supervised case of graph classification, the features of coarsened vertices are concatenated into a fully-connected layer followed by a softmax function.

\subsection{Graph Convolution}

\begin{figure*}[!t]
	\centering
	\includegraphics[width=0.9\textwidth]{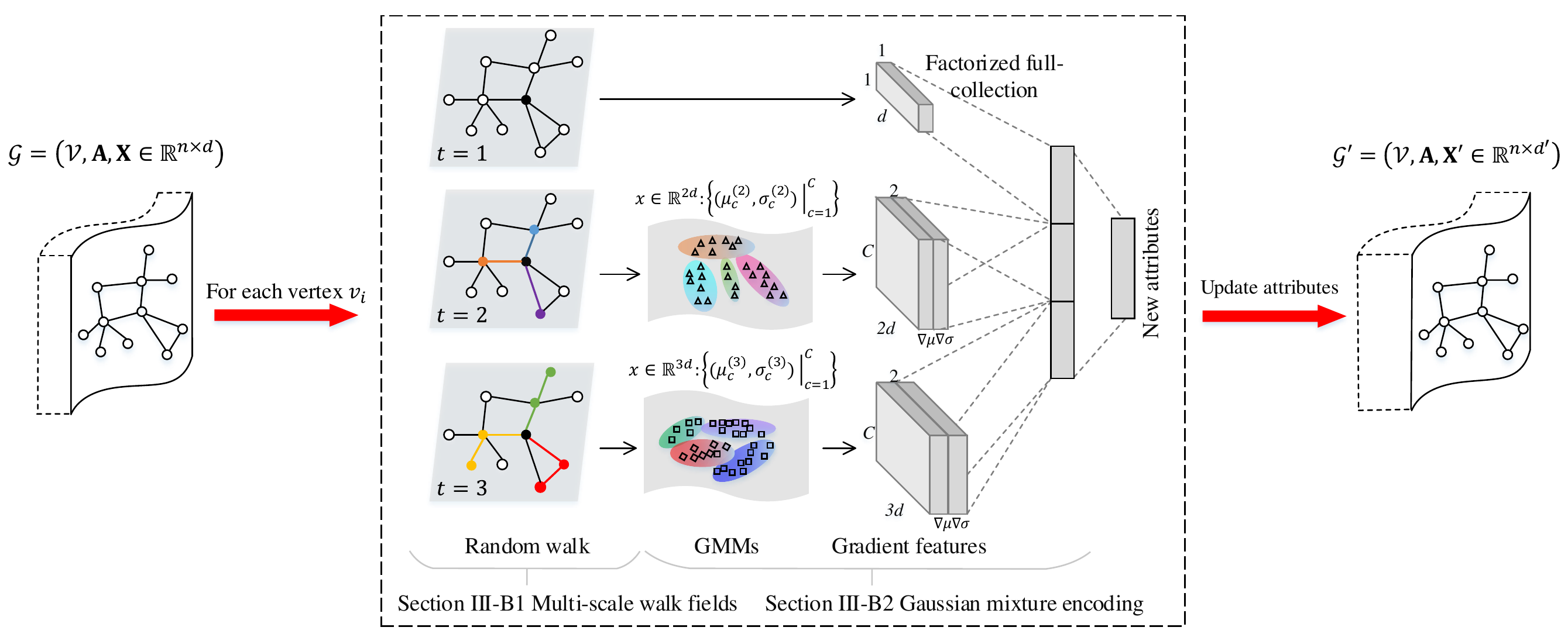}
	\caption{\Chunyan{An illustration of our graph convolution. The graph convolution layer contains a set of learnable filters, which is employed to generate a set of new attributes by filtering over outputs from the previous layer. 
	For each vertex, the graph convolution firstly defines multi-scale walk fields (Section~\ref{sec:rf}) based on random walk, which are built to characterize each local subgraph region. 
	Secondly, we design the Gaussian mixture encoding (Section~\ref{sec:gaussian}) for representing the variations of multi-scale walk fields, which is equal to graph convolution filtering. 
	\Jiang{We} then obtain gradient features of each walk field by \Jiang{Gaussian mixture encoding operation and} transform them into feature vectors with the same dimension. 
	Finally, we project all gradient features into a low \JJ{dimension} representation vector that is the new attribute of the input vertex.}}	
	\label{fig:gaussianencoding}
	\vskip -0.1in
\end{figure*}

As shown in Fig.~\ref{fig:gaussianencoding}, the graph convolution layer consists of a set of learnable filters, which are employed to generate a set of new attributes by filtering over outputs from the previous layer. The area that each filter covers is called receptive field.
Our graph convolution firstly defines multi-scale walk fields (Section~\ref{sec:rf}) based on random walk, which consist of sets of paths and vertices attributes centering on a vertex. The walk fields can cover the spatial structure and local vertices attributes. 
Secondly, we design the Gaussian mixture encoding (Section~\ref{sec:gaussian}) on walk fields, which is equal to graph convolution filtering. Considering receptive field is not related to the order of walk paths, we use \Jiang{Gaussian mixture models} to model the distribution of walk fields on \Jiang{each} scale. Moreover, Gaussian models can focus on local subgraph structure \Chunyan{due} to the fact that random walk keeps structure information of graph. 
In addition, we can obtain \Chunyan{gradient features of each walk field by \Jiang{Gaussian mixture models}} and transform them into feature vectors with the same dimension. The parameters of Gaussian models and gradient feature projections need to be learned on Gaussian mixture encoding. Finally, we project all gradient features into a low \JJ{dimension} representation vector that is the new attribute of the input vertex. The main two modules are illustrated in following sections in detail.

\subsubsection{Walk Fields}\label{sec:rf}

\begin{figure}[!t]
	\centering
	\includegraphics[width=0.3\textwidth]{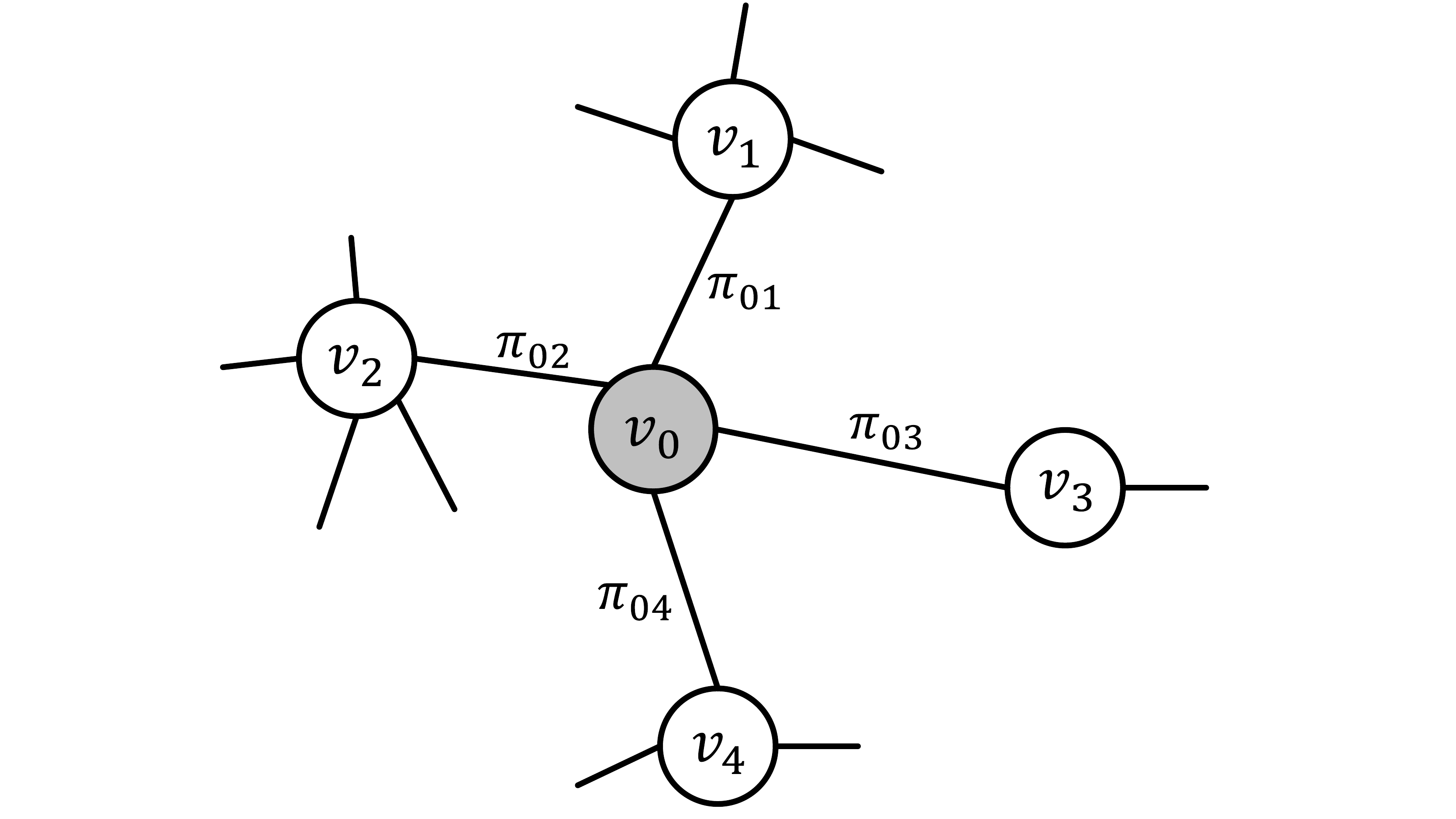}
	\caption{\Chunyan{An} illustration of the random walk procedure. The walk just \Jiang{transfers} from $v_{0}$ to  $v_{i}$ by evaluating its next step out of node $v_{0}$. Edge label $\pi_{0i}$ indicates the search probability from vertex $v_{0}$ to  $v_{i}$.}	
	\label{fig:randomwalk1}
	\vskip -0.2in
\end{figure}

\begin{figure}[!t]
	\centering
	\includegraphics[width=0.35\textwidth]{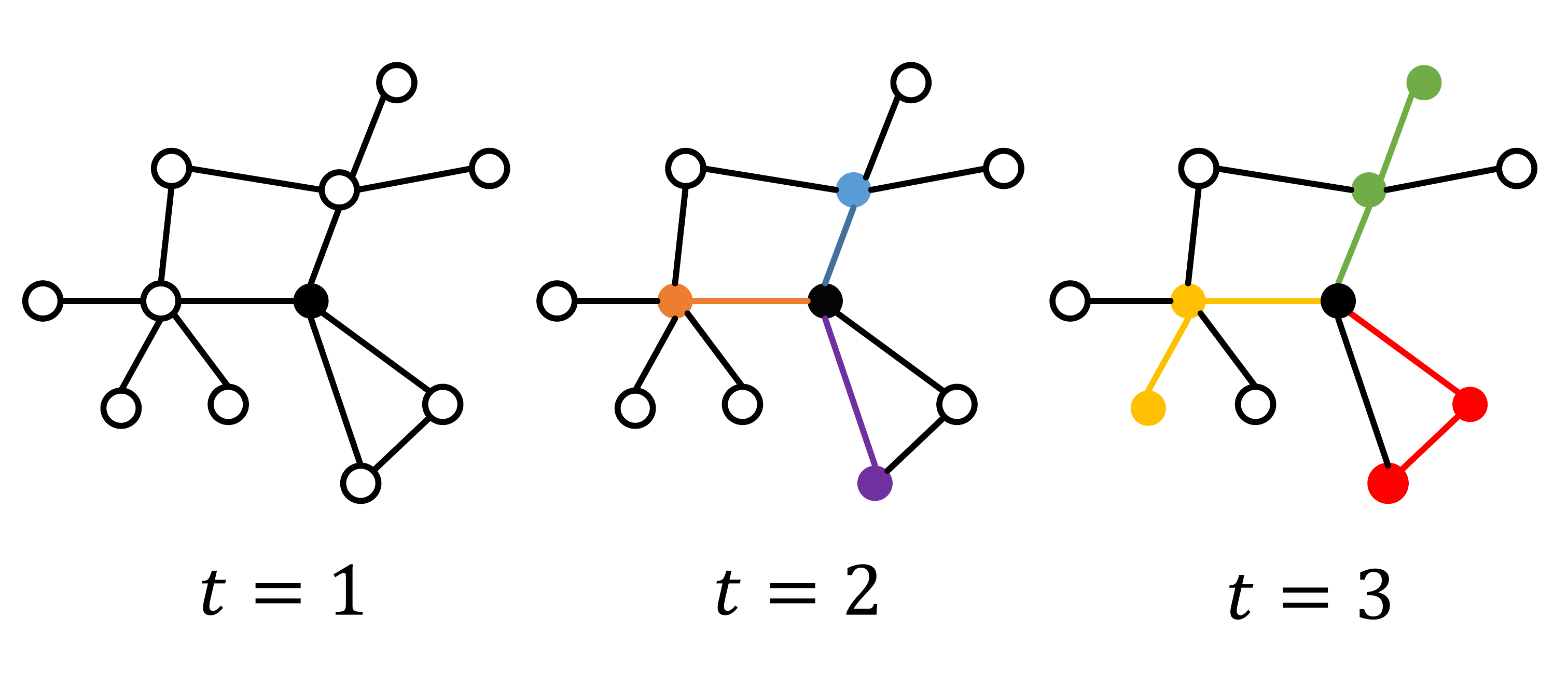}
	\caption{An illustration of our multi-scale receptive fields based on random walk. The $t$-scale walk field \Chunyan{consisting of several random walk paths} depicts the topology structure of a local subgraph within the $t$-hop neighborhood.}	
	\label{fig:randomwalk2}
	\vskip -0.1in
\end{figure}

A random walk starts at one vertex and randomly traverses next successors in depth according to the \Jiang{transition possibilities} of edge connections as shown in Fig.~\ref{fig:randomwalk1}. The walk will be stopped if the sampling length is satisfied. Formally, given the current observed vertex $v_{k}$, one neighbor vertex $v_l$ can be hit \JJ{with the probability $P(v_l|v_k)$}:
\begin{align}
P(v_l|v_k) = \left\{\begin{array}{cc}\pi_{kl}, & \text{if }(v_k,v_l)\in\mcE,\\ 0, & \text{othervise}.\end{array}\right.
\end{align}
where $\pi_{kl}$ is the transition possibility from vertices $v_k$ to $v_l$. For the adjacency matrix $\A$, those non-zero items therein indicate the connection strengthen of edges. Thus we can use the normalized adjacency matrix to randomly sample walks, \ie, the transaction matrix is defined as
\begin{equation}
\Pi = \D^{-1}\A,
\end{equation}
where $\D$ is the degree matrix of $\A$ with only non-zero diagonal elements, and $D_{ii} = \sum_{j}A_{ij}$. Thus the \Jiang{transition probability} from each vertex to all connected vertices sums to $1$. 
When we truncate the walk at the step $t$, a walk path of $t$-length is defined from the specified root vertex.

Given a reference vertex $v_i$, its $t$-scale local receptive field is actually a local subgraph consisting of its $t$-hop neighbors. 
Since random walks well preserve topology structures of graph, we define the $t$-scale receptive field as a set of $t$-length walk paths starting at the reference vertex, which we name as walk field. One advantage of walk field is \Jiang{that a subgraph is converted to} a set for well favoring the next math formulation. 
Further, multi-scale walk fields can be constructed to capture different-size receptive fields, as shown in Fig.~\ref{fig:randomwalk2}. 
We sample $K$ random walks within \Jiang{different scales of walk field $t=2,\cdots, T$}, where $T$ is the maximum scale of walk field. 
The walk paths of the \Jiatao{$t$-scale} walk field form a path set $\mcW^{v_i}_t=\{\mcP^{v_i}_{t,k} =(v_{k_0}=v_i,v_{k_1},\cdots,v_{k_t})| k=1,\cdots, K\}$, where $\mcP^{v_i}_{t,k}$ denotes the $k$-th \Jiang{walk path} of \Jiatao{$t$-length} starting at the root vertex $v_i$.

\subsubsection{Gaussian Mixture Encoding}\label{sec:gaussian}

According to the definition of walk field, given a vertex $v_i$, we can obtain a walk set $\mcW^{v_i}_t=\{\mcP^{v_i}_{t,1},\cdots, \mcP^{v_i}_{t,K}\}$ of $K$ \Jiatao{$t$-length} walks, which characterizes the corresponding $t$-hop \JJ{neighbor} relationships. 
Moreover, one walk path $\mcP^{v_i}_{t,k}$ is associated with an attribute matrix $\X_{\mcP^{v_i}_{t,k}}=[\X_{v_{k_0}},\cdots,\X_{v_{k_t}}]$, which is the attribute concatenation of the vertices in the \Jiang{walk path}. 
Thereby, the walk field of \Jiatao{$t$-scale} anchored at $v_i$ may be formulated as a walk-associated attribute set $\{\X_{\mcP^{v_i}_{t,1}},\cdots,\X_{\mcP^{v_i}_{t,K}}\}$. Now we need to encode the attribute set such that homomorphic subgraphs can produce the same response. A simple solution is to average all responses of \Jiang{walk paths}, \ie,
\begin{align}
\hbx^{v_i}_{t} = \frac{1}{K}\sum_{k=1}^K f(\X_{\mcP^{v_i}_{t,k}}),
\end{align}
where $f$ is a filtering function shared for all $K$ \Jiang{walk paths} and $\hbx^{v_i}_{t}$ is the output response of the \Jiatao{$t$-scale} walk field at vertex $v_i$. Obviously, this strategy only uses the one-order statistics (\ie, mean) without any arrangement on these \Jiang{walk paths}.

In the standard \JJ{convolution} kernel used on image, however, it is important to encode the variations of a local receptive field region. For example, let $\mcK\in\mbR^{k\times k\times c}$ denote a $k\times k$ \JJ{convolution} kernel imposed on image, where $c$ is the number of filters (or output channels). 
The $c$ filters of kernel size $k\times k$ can well encode local variations of image. 
An important reason is that the kernel is position-related within a receptive field, in which each pixel position is assigned to \Jiang{a particular mapping}. 
More intuitively, the $k^2$ pixel positions have different mappings of each other. 
Analogically, the walks within a walk field should also be associated with different filters in order to encode their internal variations (\ie, variations of walk field). However, as an unordered set, walk field does not have explicit ``directions", which distinguishes from shape-gridded images.

To address this problem, we introduce Gaussian mixture models (GMMs) to coordinate walk field. Each Gaussian model virtually defines one ``direction" of \Jiang{walk paths} therein. 
Mathematically, Gaussian models describe principal components of \Jiang{path distribution}. For the $t$-scale walk field around each vertex $v_i$, we can derive the probability estimation of each walk \Jiang{path} $\mcP^{v_i}_{t,k}$ with $C$-component GMMs,
\begin{equation}
    \label{eqn:GMM_pi}
	\begin{aligned}
	&p(\mcP^{v_i}_{t,k};\bTheta_t) = \sum_{c=1}^{C}w_{t,c}\mcN(\X_{\mcP^{v_i}_{t,k}}; \bmu_{t,c}, \bSigma_{t,c}), \\
	& \quad \quad \quad \quad \st, w_{t,c} >0, \sum_{c=1}^{C}w_{t,c}=1,
	\end{aligned}
\end{equation}
where $\bTheta_t=\{w_{t,1},\cdots,w_{t,C}, \bmu_{t,1},\cdots,\bmu_{t,C},\bSigma_{t,1},\cdots,\bSigma_{t,C}\}$ are  mixture parameters to be learned, $\{\bmu_{t,c}, \bSigma_{t,c}\}$ denotes the mean and covariance of the $c$-th Gaussian component, and $\{w_{t,c}\}$ is the corresponding weight of each Gaussian model. The Gaussian parameters are shared cross different reference vertices for walk fields with the same scale, but different on multi-scale walk fields.

In what follows, we assume attributes/signals are independent of each other, which is widely-used in signal processing. 
It means that the covariance matrix $\bSigma_{t,c}$ is diagonal, and we denote it as $\diag(\bsigma_{t,c}^2)$. To further remove the explicit constraints on $\{w_{t,c}\}$, we adopt the \Jiang{softmax} normalization by re-parameterizing them as the variables $\alpha_{t,c}$,
\begin{equation}
\begin{aligned}
w_{t,c} = \frac{\exp(\alpha_{t,c})}{\sum_{i=1}^{C}\exp(\alpha_{t,i})}.
\end{aligned}
\end{equation}
The log-likelihood of a $t$-scale walk field can be written as,
\begin{equation}
\label{eqn:GMM_log}
\begin{aligned}
\zeta(\mcW^{v_i}_{t}) &= \sum_{k=1}^{K}\ln \sum_{c=1}^{C}w_{t,c}\mcN(\X_{\mcP^{v_i}_{t,k}}; \bmu_{t,c}, \bSigma_{t,c}), \\
& = \sum_{k=1}^{K}\ln \sum_{c=1}^{C} \frac{w_{t,c}}{\eta_{t,c}}\exp{\{||-\frac{\X_{\mcP^{v_i}_{t,k}} - u_{t,c}}{\sigma_{t,c}}||^2\}},
\end{aligned}
\end{equation}
where $\eta_{t,c} = (2\pi)^{d/2}|\diag(\sigma_{t,c})|)$.

To conveniently infer forward, instead of the expectation-maximization (EM) algorithm, we derive the gradients of the log-likelihood \Chunyan{(i.e., Eq.(6))} with regard to the learned parameters $\bTheta_t=\{\balpha_{t,1}, \cdots, \balpha_{t,C}, \bmu_{t,1}, \cdots, \bmu_{t,C}, \bsigma_{t,1}, \cdots, \bsigma_{t,C}\}$, motivated by \Jiatao{the related work~\cite{sanchez2013image}}. For clarity, we simplify some notations as following:
\begin{equation}
\begin{aligned}
\mcN_{t,k,c}^{v_i} &= \mcN(\X_{\mcP^{v_i}_{t,k}}, \bmu_{t,c}, \bsigma_{t,c}^2), \\
\mcQ_{t,k,c}^{v_i} &= \frac{w_{t,c}\mcN_{t,k,c}^{v_i} }{\sum_{c=1}^{C}w_{t,c}\mcN_{t,k,c}^{v_i}}.
\end{aligned}
\end{equation}
Hence, the optimization of log-likelihood \Chunyan{(i.e., Eq.(6))} can be simply denoted as following:
\begin{equation}
\arg\max_{\bTheta_{t}} \zeta(\mcW^{v_i}_{t}) = \sum_{k=1}^{K}\ln \sum_{c=1}^{C}w_{t,c}\mcN_{t,k,c}^{v_i}.
\end{equation}
Here, we can solve the optimization problem. The gradients of Gaussian model parameters $(\balpha_{t,c}, \bmu_{t,c}, \bsigma_{t,c})$ are calculated as following:
\begin{align}
\frac{\partial \zeta(\mcW^{v_i}_{t})}{\partial \balpha_{t,c}} &= \sum_{k=1}^{K} \frac{1}{\sum_{j=1}^{C}w_{t,j}{\mcN_{t,k,j}^{v_i}}}\cdot \nonumber\\ &\left(\frac{\partial w_{t,c}\mcN_{t,k,c}^{v_i}}{\partial \balpha_{t,c}} + \frac{\sum_{j=1,j\neq c}^{C}\partial w_{t,j}\mcN_{t,k,j}^{v_i}} {\partial \balpha_{t,c}}\right), \nonumber\\
&= \sum_{k=1}^{K} \frac{1}{\sum_{j=1}^{C}w_{t,j}{\mcN_{t,k,j}^{v_i}}}\cdot \nonumber\\ &\left((1-w_{t,c})w_{t,c}\mcN_{t,k,c}^{v_i} - \sum_{j=1,j\neq c}^{C}w_{t,c}w_{t,j}\mcN_{t,k,j}^{v_i}\right), \nonumber\\
& = \sum_{k=1}^{K}\frac{w_{t,c}(\mcN_{t,k,c}^{v_i}-\sum_{j=1}^{C}w_{t,j}\mcN_{t,k,j}^{v_i})}{\sum_{j=1}^{C}w_{t,j}{\mcN_{t,k,j}^{v_i}}}, \nonumber\\
& = \sum_{k=1}^{K}\left(\frac{w_{t,c}\mcN_{t,k,c}^{v_i}}{\sum_{j=1}^{C}w_{t,j}{\mcN_{t,k,j}^{v_i}}} - w_{t,c}\right), \nonumber\\
& = \sum_{k=1}^{K}\left(\mcQ_{t,k,c}^{v_i} - w_{t,c}\right),  \nonumber \\
& = \sum_{k=1}^{K}\left(\mcQ_{t,k,c}^{v_i} - \frac{\exp{(\balpha_{t,c})}}{\sum_{i=1}^{C}\exp{(\balpha_{t,c})}}\right), \\
\frac{\partial \zeta(\mcW^{v_i}_{t})}{\partial \bmu_{t,c}} &= \sum_{k=1}^{K}\frac{w_{t,c}\mcN_{t,k,c}^{v_i}}{\sum_{j=1}^{C}w_{t,j}{\mcN_{t,k,j}^{v_i}}}\cdot\frac{\sum_{j=1}^{C}\partial\mcN_{t,k,j}^{v_i}}{\partial \bmu_{t,c}}, \nonumber\\
& = \sum_{k=1}^{K}\frac{w_{t,c}\mcN_{t,k,c}^{v_i}}{\sum_{j=1}^{C}w_{t,j}{\mcN_{t,k,j}^{v_i}}} \left(\frac{\X_{\mcP_{t,k}^{v_i}}-\bmu_{t,c}}{\bsigma_{t,c}^{2}}\right), \nonumber\\
& = \sum_{k=1}^{K}\frac{\mcQ_{t,k,c}^{v_i}(\X_{\mcP_{t,k}^{v_i}}-\bmu_{t,c})}{\bsigma_{t,c}^{2}}, \\
\frac{\partial \zeta(\mcW^{v_i}_{t})}{\partial \bsigma_{t,c}} &= \sum_{k=1}^{K}\frac{w_{t,c}\mcN_{t,k,c}^{v_i}}{\sum_{j=1}^{C}w_{t,j}{\mcN_{t,k,j}^{v_i}}}\cdot\frac{\sum_{j=1}^{C}\partial\mcN_{t,k,j}^{v_i}}{\partial \bsigma_{t,c}}, \nonumber\\
& = \sum_{k=1}^{K}\frac{w_{t,c}\mcN_{t,k,c}^{v_i}}{\sum_{j=1}^{C}w_{t,j}{\mcN_{t,k,j}^{v_i}}} \left(\frac{(\X_{\mcP_{t,k}^{v_i}}-\bmu_{t,c})^{2}}{\bsigma_{t,c}^{3}} - \frac{1}{\bsigma_{t,c}}\right), \nonumber\\
& = \sum_{k=1}^{K} \left(\frac{\mcQ_{t,k,c}^{v_i}(\X_{\mcP_{t,k}^{v_i}}-\bmu_{t,c})^{2}}{\bsigma_{t,c}^{3}} - \frac{\mcQ_{t,k,c}^{v_i}}{\bsigma_{t,c}}\right).
\end{align}
The gradients of all Gaussian mixture models' parameters $\bTheta_t$ are denoted as
\begin{equation}
\begin{aligned}
\F^{v_i}_t = [\mathscr{G}_{\balpha_{t,1}}^{\zeta}, \cdots, \mathscr{G}_{\balpha_{t,C}}^{\zeta},
\mathscr{G}_{\bmu_{t,1}}^{\zeta}, \cdots, \mathscr{G}_{\bmu_{t,C}}^{\zeta},
\mathscr{G}_{\bsigma_{t,1}} ^{\zeta}, \cdots, \mathscr{G}_{\bsigma_{t,C}}^{\zeta}],
\end{aligned}
\end{equation}
where $\F ^{v_i}_t \in \mbR ^{(2td+1)C}$, $\Jiang{\mathscr{G}_{\balpha_{t,c}}^{\zeta}}, \mathscr{G}_{\mu_{t,c}}^{\zeta}, \mathscr{G}_{\bsigma_{t,c}}^{\zeta} $ are expressed as:
\begin{equation}
\left\{
\begin{array}{cc}
\begin{aligned}
&\mathscr{G}_{\balpha_{t,c}}^{\zeta} = \sum_{k=1}^{K}\left(\mcQ_{t,k,c}^{v_i} - \frac{\exp{(\balpha_{t,c})}}{\sum_{i=1}^{C}\exp{(\balpha_{t,c})}}\right), \\
&\mathscr{G}_{\bmu_{t,c}}^{\zeta} = \sum_{k=1}^{K}\frac{\mcQ_{t,k,c}^{v_i}(\X_{\mcP_{t,k}^{v_i}}-\bmu_{t,c})}{\bsigma_{t,c}^{2}}, \\
&\mathscr{G}_{\bsigma_{t,c}}^{\zeta} = \sum_{k=1}^{K} \left(\frac{\mcQ_{t,k,c}^{v_i}(\X_{\mcP_{t,k}^{v_i}}-\bmu_{t,c})^{2}}{\bsigma_{t,c}^{3}} - \frac{\mcQ_{t,k,c}^{v_i}}{\bsigma_{t,c}}\right).
\end{aligned}
\end{array}
\right.
\end{equation}
For each reference vertex, finally, the gradient features cross all scales may be aggregated as \Jiatao{$[\X_{v_i}, \F^{v_i}_2, \cdots, \F^{v_i}_T]$}, where the first item corresponds to the vertex itself\Jiang{.} \Jiang{Moreover}, the dimension of the aggregated features becomes proportionally large with the increase of walk scales. 
To reduce computational cost, we factorize them, and then separately transform them into a low dimension space. Thus the final convolution filtering is defined as follows:
\begin{align}
\Jiatao{\tbF_{v_i} = f([\X_{v_i}, g_2(\F^{v_i}_2), \cdots, g_T(\F^{v_i}_T)])}, \label{eqn:filtering}
\end{align}
where $g_t$ is a filtering function of the $t$-scale walk field, the filtering function $f$ is imposed on the features of all scales.
In practice, $g_t$ and $f$ denote different fully-connected layers.
The walk-steered convolution algorithm is shown in Algorithm~\ref{alg: WSC}.

\begin{algorithm}[!t]
	\caption{The Walk-Steered Convolution algorithm}
	\label{alg: WSC}
	\begin{algorithmic}[1]
		\REQUIRE {Graph $\mcG=(\mcV,\A,\X)$;  vertices $\mcV = \{v_i\}_{i=1}^{m}$;  vertices attribute $\X \in \mbR^{m \times d}$;  adjacency matrix $\A \in \mbR^{m \times m}$; non-linearity $\sigma$; random sample times $K$; random sample length $T$; \Jiang{Gaussian component $C$}; random walk function $RW(\cdot; t, k)$; differentiable Gaussian encoding \Jiang{function} $G(\cdot; C)$; feature mapping function $f(\cdot)$; feature mapping \Jiang{functions} $g_{i}(\cdot), i = 2,\cdots,T$.}
		\ENSURE {Graph $\mcG'=(\mcV,\A,\X')$; \JJ{attribute} representation $\X'_{v_i}$, for all $v_i \in \mcV$}.
		\FOR {$i = 1, \cdots, m$}
		\FOR {$t = 2, \cdots, T$}
		\FOR {$k = 1, \cdots, K$}
		\STATE $\mcP^{v_i}_{t,k} \leftarrow RW(v_i; t, k)$
		\ENDFOR
		\STATE $\mcW^{v_i}_t \leftarrow [\mcP^{v_i}_{t,1},\cdots, \mcP^{v_i}_{t,K}]$
		\STATE \Jiatao{$\X_{\mcW^{v_i}_t} \leftarrow [\X_{\mcP^{v_i}_{t,1}},\cdots,\X_{\mcP^{v_i}_{t,K}}]$}
		\STATE \Jiatao{$\F_{t}^{v_i} \leftarrow G(\X_{\mcW^{v_i}_t}; C)$}
		\STATE \Jiatao{$\tbF_{t}^{v_i} \leftarrow g_{t}(\F_{t}^{v_i})$}
		\ENDFOR
		\STATE \Jiatao{$\tbF_{v_i} \leftarrow  f([\X_{v_i},\tbF_{2}^{v_i}, \cdots, \tbF_{T}^{v_i}])$}
		\STATE \Jiatao{$\X'_{v_i} \leftarrow \sigma (\tbF_{v_i})$}
		\ENDFOR
	\end{algorithmic}
\end{algorithm}

\subsection{Graph Coarsening} \label{sec:coasen}

\begin{figure}[!t]
	\centering
	\includegraphics[width=0.5\textwidth]{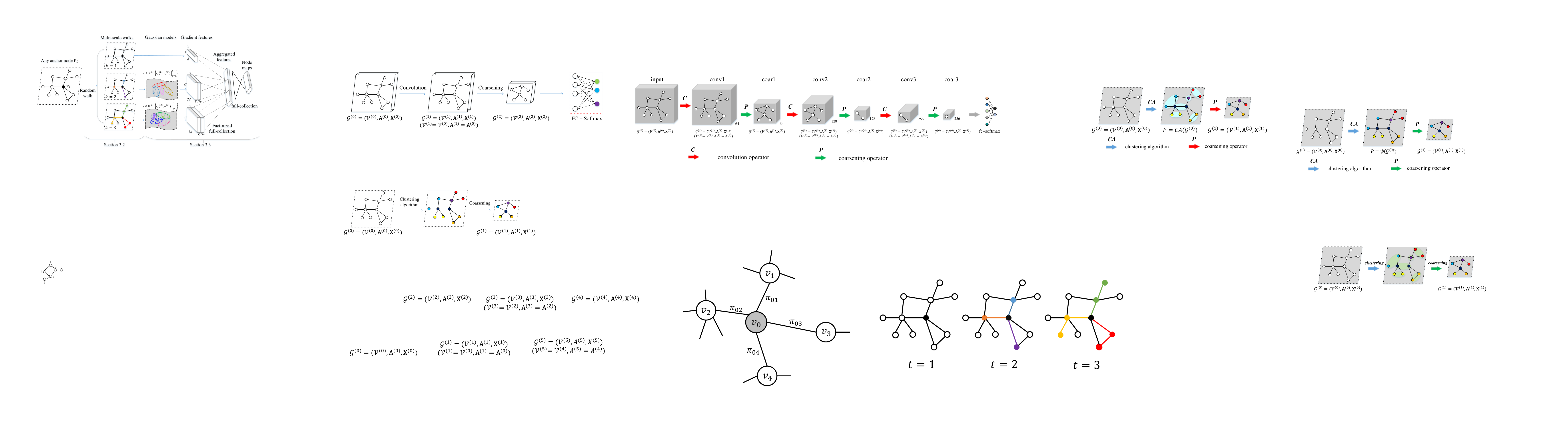}
	\caption{\Jiatao{An illustration of our graph coarsening}. 
	\Chunyan{With the input graph  $\mcG^{(0)}=(\mcV^{(0)},\A^{(0)},\X^{(0)})$, we can dynamically incorporate the information of edges by using the  clustering algorithm and then take the maximum as the final attributes of new vertex for producing a new coarsening graph $\mcG^{(1)}=(\mcV^{(1)},\A^{(1)},\X^{(1)})$. 	
	}}
	\label{fig:downsampling}
\end{figure}

CNNs usually insert a pooling layer after the convolution layer, which can extract high-level semantic information and reduce the computation complexity of network by decreasing the spatial size of convolution map.
What's more, the pooling operator can transform original input into a hierarchical representation and alleviate the overfitting problem to some extent. The pooling operator is independent on \Chunyan{each} channel of input. 
%
%
We often adopt a pooling layer with the kernel size $2\times 2$ applied with a stride of 2, which can discard 75\% of input activations.  
The maximum operation would in this case be taken a maximum value over 4 signals/activations.
For the graph convolutional neural network, we downsample some vertices through graph coarsening to abstract high-level semantics and reduce computational burden.
\Chunyan{Different from the pooling operation tailoring shape-gridded images, graph coarsening is performed on irregular topology structures.
We firstly implicitly incorporate the information of edges by using the clustering algorithm~\cite{dhillon2007weighted,jiang2018gaussian}, where each vertex can be assigned to a weight for marking its importance.
For completing the graph coarsening process, the maximum value of attributes in a cluster can be taken as the final attributes of new vertex. If any two vertices respectively from two clusters are connected, the new two vertices w.r.t. both clusters will also be connected. The illustration of graph coarsening can be seen in Fig.~\ref{fig:downsampling}. }


\Jiatao{Given a graph data as a tuple $\mcG^{(0)}=(\mcV^{(0)}, \A^{(0)}, \X^{(0)})$, 
we use the multi-layer perception (\eg, the function $\Jiang{\phi(\X^{(0)}_{v_{i}})}$) to learn the corresponding weight $\gamma_{i}$ of each vertex $v_{i}$ in our work. 
With the existing clustering algorithm~\cite{jiang2018gaussian}, we can learn the probabilities that each vertex falls into the $n$ clusters with the learning weight vector and adjacency matrix, \ie, $\P =\psi (\Gamma; \A^{(0)})$. 
By employing the maximum probability of each cluster, we can gain a binary coarsening (pooling or downsampling) matrix $\P \in \mbR^{m \times n}$, where $m$ and $n$ are the number of nodes on graphs $\mcG^{(0)}$ and $\mcG^{(1)}$. 
%
%
For any element $p_{ij} \in \P$, $p_{ij} \in \{0, 1\}$, the $i$-th row of matrix $\P$ only contains a non-zero element. The new coarsening  graph $\mcG^{(1)}=(\mcV^{(1)}, \A^{(1)}, \X^{(1)})$ can be calculated as following:

\begin{equation}
\A^{(1)} = \P^T \A^{(0)} \P, \quad \X^{(1)} = \X^{(0)} \Jiatao{\otimes} \P.
\end{equation}
Note that \Jiatao{$\otimes$} is a max-pooling operator that projects maximum values of attributes in a cluster as the final attributes of new vertex. The graph coarsening algorithm is shown in Algorithm~\ref{alg: Pooling}.
}

\begin{algorithm}[!t]
	\caption{The Graph Coarsening algorithm}
	\label{alg: Pooling}
	\begin{algorithmic}[1]
		\REQUIRE {Graph \Jiatao{$\mcG=(\mcV^{(0)},\A^{(0)},\X^{(0)})$}; \Jiang{vertices $\mcV^{(0)} = \{v_i\}_{i=1}^{m}$}; \Jiang{vertices attribute $\X^{(0)} \in \mbR^{m \times d}$}; \Jiang{adjacency matrix $\A^{(0)} \in \mbR^{m \times m}$}; differentiable weight \Jiang{function} $\phi (\cdot)$; clustering function \Jiatao{$\psi(\cdot; \A^{(0)})$}; max-pooling operator \Jiatao{$\otimes$}}.
		\ENSURE {Graph $\mcG^{(1)}=(\mcV^{(1)},\A^{(1)},\X^{(1)})$; \JJ{attribute} representation $\X^{(1)}$; \JJ{adjacency} matrix $\A^{(1)}$}.
		\FOR {$i = 1, \cdots, m$}
		\STATE \Jiatao{$\gamma_i \leftarrow \Jiang{\phi(\X^{(0)}_{v_i})}$.}
		\ENDFOR
		\STATE \Jiatao{$\Gamma \leftarrow [\gamma_{1}, \cdots, \gamma_{m}]$.}
		\STATE $\P \leftarrow \Jiatao{\psi (\Gamma; \A^{(0)})}, \P \in \mbR^{m \times n}$ is a binary matrix.
		\STATE $\A^{(1)} \leftarrow \P^T \Jiatao{\A^{(0)}} \P, \quad \X^{(1)} \leftarrow \Jiatao{\X^{(0)}} \Jiatao{\otimes} \P$.
	\end{algorithmic}
\end{algorithm}

\subsection{Loss Function}\label{sec:loss}

Here we deal with the graph classification in a supervised way. Given $n$ training examples of attribute graphs $\mcG = \{ \mcG_1, \mcG_2, \cdots, \mcG_n\}$ and their corresponding labels $\Y = \{y_1, y_2, \cdots, y_n|y_i \in \{0, 1, \cdots, l\}\}$, where $l$ is the number of class of attribute graphs. Given the model (WSC) projection \Jiatao{$\varphi(\cdot)$}, the predicted probability for $j$ class for input $\mcG_i$ is denoted,
\begin{equation}
\left\{
\begin{array}{cc}
\begin{aligned}
&x_i = \Jiatao{\varphi}(\mcG_i), \\
&\Pr(y_i = j|x_i) = \Jiang{\frac{\exp{x_{i}^{(j)}}}{\sum_{k=0}^{l}\exp{x_{i}^{(k)}}}}.
\end{aligned}
\end{array}
\right.
\end{equation}
To minimize the cross-entropy loss between the predictions and the ground-truth, the loss function is defined as following:
\begin{equation}
\mathcal{L} =  -\frac{1}{n}\sum_{i=1}^{n}\Pr(y_i|x_i).
\end{equation}
We use the stochastic gradient descent solver for learning the
parameters of the WSC network in the task of graph classification.

\section{Experiments}

\subsection{Dataset} \label{sec:datasets}
We evaluate the effectiveness of our proposed WSC on several public graph datasets including social network and bioinformatics datasets. Their statistical properties of graph data are summarized in Table~\ref{table:benchmarkdatasets}.
The average numbers of nodes and edges describe the degree information of vertices in the graph datasets.
The average and maximum number of nodes are also important properties of graph data.

\begin{table}[!t]
	\centering
	\caption{Statistics and properties for Bioinformatics and Social network datasets.}
	\setlength{\tabcolsep}{1pt}
	\scalebox{1}{
		\begin{tabular}{|l| c c c c c c|}
			\toprule
			Dataset & Num.       & Classes & Labels & Avg.nodes & Avg.edges & Max.nodes\\
			\midrule
			MUTAG   & 188        & 2       & 7             & 17           & 19   & 28       \\
			PTC     & 344        & 2       & 19            & 14           & 14   & 109      \\
			NCI1    & 4110       & 2       & 37            & 29           & 32   & 111      \\
			ENZYMES & 600        & 6       & 3             & 32           & 62   & 126      \\
			PROTEINS& 1113       & 2       & 3             & 39           & 72   & 620      \\
			\midrule
			COLLAB             & 5000    & 3             & --           & 74        & 2457   & 492       \\
			REDDIT-BINARY      & 2000    & 2             & --           & 429       & 497    & 3782      \\
			REDDIT-MULTI-5K    & 5000    & 5             & --           & 508       & 594    & 3648     \\
			REDDIT-MULTI-12K   & 11929   & 11            & --           & 391       & 456    & 3782    \\
			IMDB-BINARY        & 1000    & 2             & --           & 19        & 96     & 136       \\
			IMDB-MULTI         & 1500    & 3             & --           & 13        & 65     & 89         \\
			\bottomrule
		\end{tabular}
	}
	\label{table:benchmarkdatasets}
	\vskip -0.1in
\end{table}

\textbf{Bioinformatics datasets}: MUTAG~\cite{debnath1991structure} consists of 188 chemical compounds with 7 discrete node labels. PTC~\cite{toivonen2003statistical} contains 344 chemical compounds with 19 discrete node labels. NCI1~\cite{wale2008comparison}, which is collected by National Cancer Institute (NCI), contains a balanced subset of 4110 chemical compounds with 37 discrete node labels about the human tumor cell. ENZYMES~\cite{borgwardt2005protein} consists of 600 protein tertiary structures with 3 discrete node labels, which are equally divided into 6 classes. PROTEINS~\cite{borgwardt2005protein} contains 1113 protein structures of secondary structure elements (SSEs) with 3 discrete node labels.

\textbf{Social network datasets}: COLLAB~\cite{leskovec2005graphs} is a scientific collaboration dataset. This is mainly acquired from 3 collaboration datasets including \textit{High Energy Physics}, \textit{Condense Mater Physics} and \textit{Astro Physics}. COLLAB contains 5000 ego-collaboration graphs with three classes. REDDIT-BINARY, REDDIT-MULTI-5K, REDDIT-MULTI-12K are three balanced datasets with two, five and eleven classes respectively. They consist of a series of graphs that correspond to online discussion threads, where nodes denote users and edges are built between two nodes if at least one of them responds to another's comment. IMDB-BINARY and IMDB-MULTI belong to movie collaboration datasets derived from IMDB with two and three classes respectively, where nodes denote actors/actresses and edges are built between them if they have appeared in the same movie.

\subsection{Implementation Details} \label{sec:details}

\begin{table}[!t]
	\centering
	\caption{The detailed configuration of WSC network and the feature size of each layer. \Chunyan{$m, d$ and $c$ denote the number of vertices, the attribute dimension of each vertex and the number of categories, respectively.}}
	\setlength{\tabcolsep}{4pt}
	\scalebox{1}{
		\begin{tabular}{|l |c |c |c |c|}
			\toprule
			Schema & Filter & Filter size & Attribute size & vertex size \\
			\midrule
			Input                  & & &$m \times d$ &$m$ \\
			\midrule
			Convolution layer 1    &64    &$d \times 64$    &$m \times 64$     &$m$ \\
			\midrule
			Coarsening layer 1     &  &  &$m/4 \times 64$    &$m/4$ \\
			\midrule
			Convolution layer 2    &128  &$64 \times 128$    &$m/4 \times 128$  &$m/4$  \\
			\midrule
			Coarsening layer 2     &  &  &$m/16 \times 128$    &$m/16$ \\
			\midrule
			Convolution layer 3    &256    &$128 \times 256$     &$m/16 \times 256$     &$m/16$ \\
			\midrule
			Coarsening layer 3     &  &  &$1 \times 256$    &$1$ \\
			\midrule
			Fully connected layer  & & & $1 \times 256$  & \\
			\midrule
			Output     & & &c & \\
			\bottomrule
		\end{tabular}
	}
	\label{table:config}
	\vskip -0.1in
\end{table}

In default, our model of WSC consists of three convolution layers, each of which follows by a graph coarsening layer, and a fully connected layer with a softmax loss layer. 
Since the MUTAG dataset has less vertices, a simple network structure with 5 layers (i.e., N=5) can be described as C(64)-P(0.25)-C(128)-P(0.0)-FC(256), where C, P and FC denote convolution, coarsening and fully connected layers, respectively. 
The values in brackets mean the channel number, the coarsening ratio or the neuron number. 
P(0.0) means that the coarsening operator downsamples all vertices to only one vertex.
On all other graph datasets, the network configuration with 7 layers (i.e., N=7) is described as C(64)-P(0.25)-C(128)-P(0.25)-C(256)-P(0.0)-FC(256). 
Table~\ref{table:config} shows the detail configuration of WSC network and the feature size of each layer, where $m$ is the number of graph vertices and $d$ is the attribute dimension of each vertex.
The scale number $T$ of walk fields and the Gaussian number $C$ are both set to 3. The sampling times $K$ of \Jiang{walk paths} can be set to 8. The fully connected layer is also followed by a dropout layer with rate 0.5. The nonlinear activate function uses ReLU unit for each convolution layer and fully connected layer. We use stochastic gradient descent to train WSC with 400 epochs, batch size of 100 graphs, learning rate of 0.1, and momentum of 0.95. The parameters analysis can be found in Section~\ref{sec:ps}.
What's more, the initial input of model is the attribute matrix and adjacency matrix of a graph. For the attribute matrix, we use vertex label and degree to initialize the attribute of each vertex in bioinformatics datasets. Meanwhile, we use the vertex degree as attributes in social network datasets. 
\Chunyan{Following previous standard protocols~\cite{niepert2016learning,yanardag2015deep}, we use ten-fold cross validation to evaluate the graph classification performance, nine-fold for training and one-fold for testing. In order to exclude random effects of the fold assignments, we repeat all experiments ten times, and then report average accuracies as well as standard deviations.}

\subsection{Comparisons with Baselines}

\begin{table}[!t]
	\centering
	\caption{\Chunyan{Comparisons of graph classification with different graph convolution methods.}}
	\setlength{\tabcolsep}{3.5pt}
	\label{table:conv}
	\begin{center}
		\scalebox{1}{
		\begin{tabular}{l c c c c}
				\toprule
			Dataset  & WSC w/ Cheb~\cite{defferrard2016convolutional}   & WSC w/ GC~\cite{kipf2016semi}    &WSC \\
			\midrule
			MUTAG          & 89.44 $\pm$ 6.30      & 92.22 $\pm$ 5.66     & \textbf{93.33 $\pm$ 4.15} \\
			PTC            & 68.23 $\pm$ 6.28      & 71.47 $\pm$ 4.75     & \textbf{75.29 $\pm$ 7.34} \\
			NCI1           & 73.96 $\pm$ 1.87      & 76.39 $\pm$ 1.08     & \textbf{81.70 $\pm$ 0.98}	 \\		
			ENZYMES        & 52.83 $\pm$ 7.34      & 51.50 $\pm$ 5.50     & \textbf{56.16 $\pm$ 5.91} \\
			PROTEINS       & 78.10 $\pm$ 3.37      & 80.09 $\pm$ 3.20     & \textbf{81.08 $\pm$ 2.41} \\
			\hline 
			COLLAB          &71.80 $\pm$ 1.28     &62.56 $\pm$ 2.255    &\textbf{83.96 $\pm$ 1.33} \\
			REDDIT-B        &\textbf{85.55 $\pm$ 2.25}     &82.70 $\pm$ 2.14     &84.00 $\pm$ 2.08 \\
			REDDIT-5K       &50.98 $\pm$ 1.53     &47.11 $\pm$ 2.37     &\textbf{52.40 $\pm$ 1.72}	 \\		
			REDDIT-12K      &38.70 $\pm$ 1.12     &32.17 $\pm$ 2.01     &\textbf{42.42 $\pm$ 0.61} \\
			IMDB-B          &77.00 $\pm$ 3.09     &76.40 $\pm$ 4.34     &\textbf{78.40 $\pm$ 3.95} \\
			IMDB-M          &52.80 $\pm$ 2.82     &52.53 $\pm$ 3.27     &\textbf{55.00 $\pm$ 4.51} \\						
			\bottomrule
		\end{tabular}
		}
	\end{center}
	\vskip -0.1in
\end{table}


\begin{table}[!t]
	\centering
	\caption{Comparisons of graph classification with/without our proposed convolution and coarsening operators. \Chunyan{The ``C" and ``P" denote the graph convolution and graph coarsening, respectively.} }
	\setlength{\tabcolsep}{5pt}
	\label{table:pooling}
	\begin{center}
	\scalebox{1}{
		\begin{tabular}{l c c c c}
			\toprule
			Dataset & WSC w/o P\& C & WSC w/o P& WSC \\
			\midrule
			MUTAG          & 88.66 $\pm$ 8.31      & 91.66 $\pm$ 5.12     & \textbf{93.33 $\pm$ 4.15} \\
			PTC            & 64.11 $\pm$ 6.55      & 60.29 $\pm$ 9.13     & \textbf{75.29 $\pm$ 7.34} \\
			NCI1           & 71.82 $\pm$ 1.85      & 71.84 $\pm$ 7.79     & \textbf{81.70 $\pm$ 0.98}	 \\		
			ENZYMES        & 42.33 $\pm$ 4.22      & 21.16 $\pm$ 3.57     & \textbf{56.16 $\pm$ 5.91} \\
			PROTEINS       & 77.38 $\pm$ 2.97      & 59.63 $\pm$ 4.91     & \textbf{81.08 $\pm$ 2.41} \\
			\bottomrule
		\end{tabular}
	}
	\end{center}
	\vskip -0.2in
\end{table}

\begin{table*}[!t]
	\centering
	\caption{Comparisons of graph classification with state-of-the-art methods.}
	\setlength{\tabcolsep}{3.5pt}
	\begin{center}
	\scalebox{1}{
		\begin{tabular}{|l| c c c| c c | c c c |c |c |c |c|}
			\toprule
			\multirow{2}{*}{Dataset}
			&\multicolumn{3}{|c|}{graph convolution} & \multicolumn{2}{|c|}{feature based}  & \multicolumn{3}{|c|}{graph kernel}  & \multicolumn{1}{|c|}{random walk} & \multicolumn{1}{|c|}{NN} &\multirow{2}{*}{WSC}\\
			\cline{2-11}
			&\Jiang{PSCN}~\cite{niepert2016learning}  &DCNN~\cite{atwood2016diffusion}  &NgramCNN~\cite{luo2017deep}   &FB~\cite{barnett2016feature}  &DyF~\cite{gomez2017dynamics}   &WL~\cite{morris2017glocalized}   &GK~\cite{shervashidze2009efficient}  &DGK~\cite{yanardag2015deep}  &RW~\cite{gartner2003graph}   &SAEN~\cite{orsini2018shift}  & \\
			
			\midrule
			\multirow{2}{*}{MUTAG}
			&92.63   &66.98  &\textbf{94.99}  &84.66   &88.00    &78.3   &81.66     &82.66      &83.72     &84.99 &93.33  \\		
			& $\pm$4.21  &--    &\textbf{$\pm$5.63} &$\pm$2.01 & $\pm$2.37  & $\pm$1.9 & $\pm$2.11  & $\pm$1.45  & $\pm$1.50   & $\pm$1.82  	&$\pm$4.15      \\
			
			\midrule
			\multirow{2}{*}{PTC}
			&60.00   &56.60  &68.57  &55.58   &57.15     &--     &57.26     &57.32   &57.85    &57.04  &\textbf{75.29}   \\			
			& $\pm$4.82  &--   &$\pm$1.72  &2.30   & $\pm$1.47  & --  &  $\pm$1.41  &  $\pm$1.13  & $\pm$1.30
			&  $\pm$ 1.30  & \textbf{$\pm$ 7.34}      \\
			
			\midrule
			\multirow{2}{*}{NCI1}
			&78.59   &62.61     &--  &62.90   &68.27  &\textbf{83.1}   &62.28   &62.48   &48.15    &77.80  &81.70    \\			
			& $\pm$1.89  &--   &--  &$\pm$0.96  & $\pm$0.34  &\textbf{$\pm$0.2}  & $\pm$0.29  &  $\pm$0.25  &  $\pm$0.50  & $\pm$ 0.42  & $\pm$0.98    \\
			
			
			\midrule
			\multirow{2}{*}{ENZYMES}
			& --    &18.10    &--  &29.00   & 33.21   & 53.4   & 26.61   & 27.08   & 24.16   & --   & \textbf{56.16}  \\			
			& --    &--    &--  &$\pm$1.16  & $\pm$ 1.20  & $\pm$ 1.4   &  $\pm$ 0.99  & $\pm$ 0.79  &  $\pm$ 1.64
			&--  & \textbf{$\pm$ 5.91}     \\
			
			\midrule
			\multirow{2}{*}{PROTEINS}
			& 75.89  &--   &75.96  &69.97 & 75.04   & 73.7  & 71.67   & 71.68  & 74.22    & 75.31  & \textbf{81.08}     \\			
			& $\pm$ 2.76  &--   &$\pm$2.98  &$\pm$1.34  & $\pm$ 0.65  &  $\pm$ 0.5  & $\pm$ 0.55  &  $\pm$ 0.50  & $\pm$ 0.42	&  $\pm$ 0.70 & \textbf{ $\pm$ 2.41}      \\
			
			\midrule
			\multirow{2}{*}{COLLAB}
			& 72.60  &--    &--  &76.35 & 80.61 & -- & 72.84 & 73.09 & 69.01 & 75.63 & \textbf{83.96}     \\			
			& $\pm$ 2.15  &--   &--  &1.64  & $\pm$ 1.60  & --  & $\pm$ 0.28  & $\pm$ 0.25  & $\pm$ 0.09
			& $\pm$ 0.31  & \textbf{$\pm$ 1.33}      \\
			
			\midrule
			\multirow{2}{*}{REDDIT-B}
			& 86.30  &--     &--  &88.98 &\textbf{89.51}  &75.3 & 77.34 & 78.04 & 67.63  & 86.08 & 84.00   \\			
			& $\pm$ 1.58  &--   &--  &$\pm$2.26  & \textbf{$\pm$ 1.96}  & $\pm$ 0.3  & $\pm$ 0.18  & $\pm$ 0.39  & $\pm$ 1.01
			& $\pm$ 0.53  & $\pm$ 2.08    \\
			
			\midrule
			\multirow{2}{*}{REDDIT-5K}
			& 49.10  &--      &--  &50.83   & 50.31 & --  & 41.01 & 41.27 & --  &52.24 &\textbf{52.40} \\		
			& $\pm$ 0.70  &--    &--  &1.83  &$\pm$ 1.92  & --  & $\pm$ 0.17  & $\pm$ 0.18  & --
			& $\pm$ 0.38  &\textbf{$\pm$ 1.72}   \\
			
			\midrule
			\multirow{2}{*}{REDDIT-12K}
			& 41.32  &--      &--  &42.37  & 40.30 & --  & 31.82 & 32.22 & --  & \textbf{46.72} & 42.42 \\			
			& $\pm$ 0.42  &--    &--  &1.27   & $\pm$ 1.41  & -- & $\pm$ 0.08  & $\pm$ 0.10  & --
			& \textbf{$\pm$ 0.23} & $\pm$ 0.61    \\
			
			\midrule
			\multirow{2}{*}{IMDB-B}
			& 71.00   &--    &71.66  &72.02 & 72.87 & 72.4 & 65.87 & 66.96 & 64.54 	& 71.26 & \textbf{78.40}     \\			
			& $\pm$ 2.29  &--    &$\pm$2.71  &$\pm$4.71  & $\pm$ 4.05  & $\pm$ 0.5  & $\pm$ 0.98  & $\pm$ 0.56  & $\pm$ 1.22
			& $\pm$ 0.74  & \textbf{$\pm$ 3.95}      \\
			
			\midrule
			\multirow{2}{*}{IMDB-M}
			& 45.23 &--     &50.66  &47.34  & 48.12 & -- & 43.89 & 44.55 & 34.54 & 49.11 & \textbf{55.00}      \\			
			& $\pm$ 2.84  &--  &$\pm$4.10  &3.56  & $\pm$ 3.56  & -- & $\pm$ 0.38  & $\pm$ 0.52  & $\pm$ 0.76
			& $\pm$ 0.64  & \textbf{$\pm$ 4.51}      \\
			\bottomrule
		\end{tabular}
	}
	\end{center}
	\label{table:state-of-the-art}
	\vskip -0.1in
\end{table*}

To compare our proposed graph convolution with other existing graph convolutions, we use the same network settings with our WSC, but with different graph convolutions (including ChebNet~\cite{defferrard2016convolutional} and GCN~\cite{kipf2016semi} methods), named ``WSC w/ CHEB" and ``WSC w/ GC", respectively.
\Chunyan{Table~\ref{table:conv} has shown comparisons of graph classification results with different graph convolution methods on social network and bioinformatics datasets.}
Further, we design two graph baseline networks to verify the efficacy of our proposed convolution and coarsening operators, and the corresponding comparisons of graph classification can be shown in  Table~\ref{table:pooling}.
``WSC w/o P \& C" is without our proposed convolution and coarsening operators, which can be instead of a fully connected layer.
"WSC w/o P" does not use the proposed coarsening operator in the graph network.
%
We add some blank vertices corresponding to different size graph on the experiments. The results are shown in Table~\ref{table:pooling}. From these experimental results, we have the following observations:
\begin{itemize}
	\item \Chunyan{Compared with different graph convolution methods, our graph convolution method is an effective way to aggregate the local information for learning graph representation. The WSC has a better performance than convolution methods of ChebNet~\cite{defferrard2016convolutional} and GCN~\cite{kipf2016semi} on social network and bioinformatics datasets, except on REDDIT-B dataset. When compared with graph convolution method of GCN, we obtain a large gain about 5.31\% on NCI1 dataset and 2.47\% on IMDB-M dataset. What's more, GCN~\cite{kipf2016semi} is an approximation of ChebNet~\cite{defferrard2016convolutional} method. They have a similar performance on bioinformatics datasets, because they have the similar receptive field and complexity with many stacked convolution layers. The efficacy of our graph convolution operator mainly benefits from sufficiently considering the local structure of subgraph by Gaussian encoding.}
	
	\item Compared with our WSC without graph coarsening layer (``WSC w/o P"), ours have a significant gain on bioinformatics datasets. Especially, ours have big improvements about 15.00\% and 9.96\% on PTC and NCI1 datasets. The WSC without graph coarsening layer easily overfits the sample data and the blank vertices can bring up some noises to change the original data distribution. Moreover, experimental results without convolution and coarsening layers (``WSC w/o P \& C") verify the ability of convolution and coarsening operators, which can alleviate the overfitting problem of graph network to some extent.
	
	\item Compared between convolution and coarsening operators, convolution operator can efficiently aggregate the local information and coarsening operator can efficiently prevent overfitting of model and reduce parameters of model. The WSC of stacked convolution and coarsening layers can learn a very efficient graph representation for graph-structured data, and they can be easily extended to other graph networks.
	
\end{itemize}

\subsection{Comparisons with the State-of-the-arts}

We compare our WSC with several state-of-the-arts, including recent graph convolution methods (PSCN~\cite{niepert2016learning}, DCNN~\cite{atwood2016diffusion}, NgramCNN~\cite{luo2017deep}), feature based algorithms (DyF~\cite{gomez2017dynamics}, FB~\cite{barnett2016feature}), graph kernel approaches (GK~\cite{shervashidze2009efficient}, DGK~\cite{yanardag2015deep}, WL~\cite{morris2017glocalized}), random walk based methods (RW~\cite{gartner2003graph}), and neural network methods (SAEN~\cite{orsini2018shift}). The experimental results are summarized in Table~\ref{table:state-of-the-art}, where most of them come from the related literatures. From these results, we have the following observations.
\begin{itemize}
	\item In those conventional methods, graph kernel is an effective way to learn graph representation. WL has the best performance on NCI1. Recently, the feature-based approaches (DyF~\cite{gomez2017dynamics}, FB~\cite{barnett2016feature}) endeavor to integrate multiplex features to boost classification accuracies. For example, DyF~\cite{gomez2017dynamics} uses more efficient multiplex features to discriminate different classes, and obtains a better performance on REDDIT-BINARY.
	
	\item Deep learning based methods on graph-structured data (including DCNN~\cite{atwood2016diffusion}, PSCN~\cite{niepert2016learning}, NgramCNN~\cite{luo2017deep}, SAEN~\cite{orsini2018shift} and WSC) are superior to those conventional methods in most cases. The conventional kernel methods usually require the calculation on graph kernels with the high-computational complexity. In contrast, these graph neural networks attempt to learn more  high-level features by performing inference-forward, which need relative low computational cost.
	
	\item Compared with recent graph convolutional neural networks (e.g., DCNN~\cite{atwood2016diffusion}, PSCN~\cite{niepert2016learning}, and NgramCNN~\cite{luo2017deep}), our WSC can achieve better performances on most graph datasets. On IMDB-BINARY and IMDB-MULTI datasets, ours can have the improvements of about 5\%. The main reason should be that local variations of receptive field regions are accurately described with Gaussian component analysis. Besides, walk fields can preserve local topologies well for Gaussian encoding. Compared with these kernel methods (\eg, GK~\cite{shervashidze2009efficient}, DGK~\cite{yanardag2015deep}, WL~\cite{morris2017glocalized}), our WSC has a significant improvement on most datasets except NCI1. For example, our WSC obtains a large gain on MUTAG dataset, e.g., 15.03\%, 11.67\%, 10.68\% w.r.t. WL~\cite{morris2017glocalized}, GK~\cite{shervashidze2009efficient}, DGK~\cite{yanardag2015deep}.
	
	\item Our WSC achieves state-of-the-art results on most graph datasets.
	For MUTAG, NCI1, REDDIT-BINARY and REDDIT-MULTI-12K datasets, our WSC has a completely comparable performance compared with NgramCNN~\cite{luo2017deep}, WL~\cite{morris2017glocalized}, DyF~\cite{gomez2017dynamics} and SEAN~\cite{orsini2018shift} methods, which have also obtained good performances.
	The best performances have been achieved on several bioinformatics datasets and social network datasets including PTC, ENZYMES, PROTEINS, COLLAB, IMDB-BINARY, IMDB-MULTI, and REDDIT-MULTI-5K. For example, our WSC can significantly outperform several baselines: 15.29\% over PSAN~\cite{niepert2016learning} on PTC dataset, 10.9\% over DGK~\cite{yanardag2015deep} on COLLAB dataset, 6.38\% over FB~\cite{barnett2016feature} on IMDB-B dataset.
\end{itemize}

\subsection{\Chunyan{Algorithm Analysis}} \label{sec:ps}

We discuss the impacts of three main parameters: walk field scale $T$, \Chunyan{the number of Gaussian components $C$}, the number of network layers $N$ \Chunyan{and the sampling times K of walks}. Actually, \Chunyan{these four parameters are intertwined each other to determine the final graph classification performance.} For example, increasing walk field scale $T$ and stacking multiple convolutional layers can both enlarge receptive field size, which usually results into better performance as demonstrated in conventional CNNs. Besides, Gaussian component number $C$ should be set to be proportional to walk filed scale $T$, as large receptive fields usually contain more complicated variations. Below we do an ablation study on these parameters.

\begin{figure}[!t]
	\begin{center}
		\centering
		\subfigure[\Chunyan{With variable $C$ and fixed $T=9$}]{
			\label{fig:ct_c} 
			\includegraphics[width=0.23\textwidth]{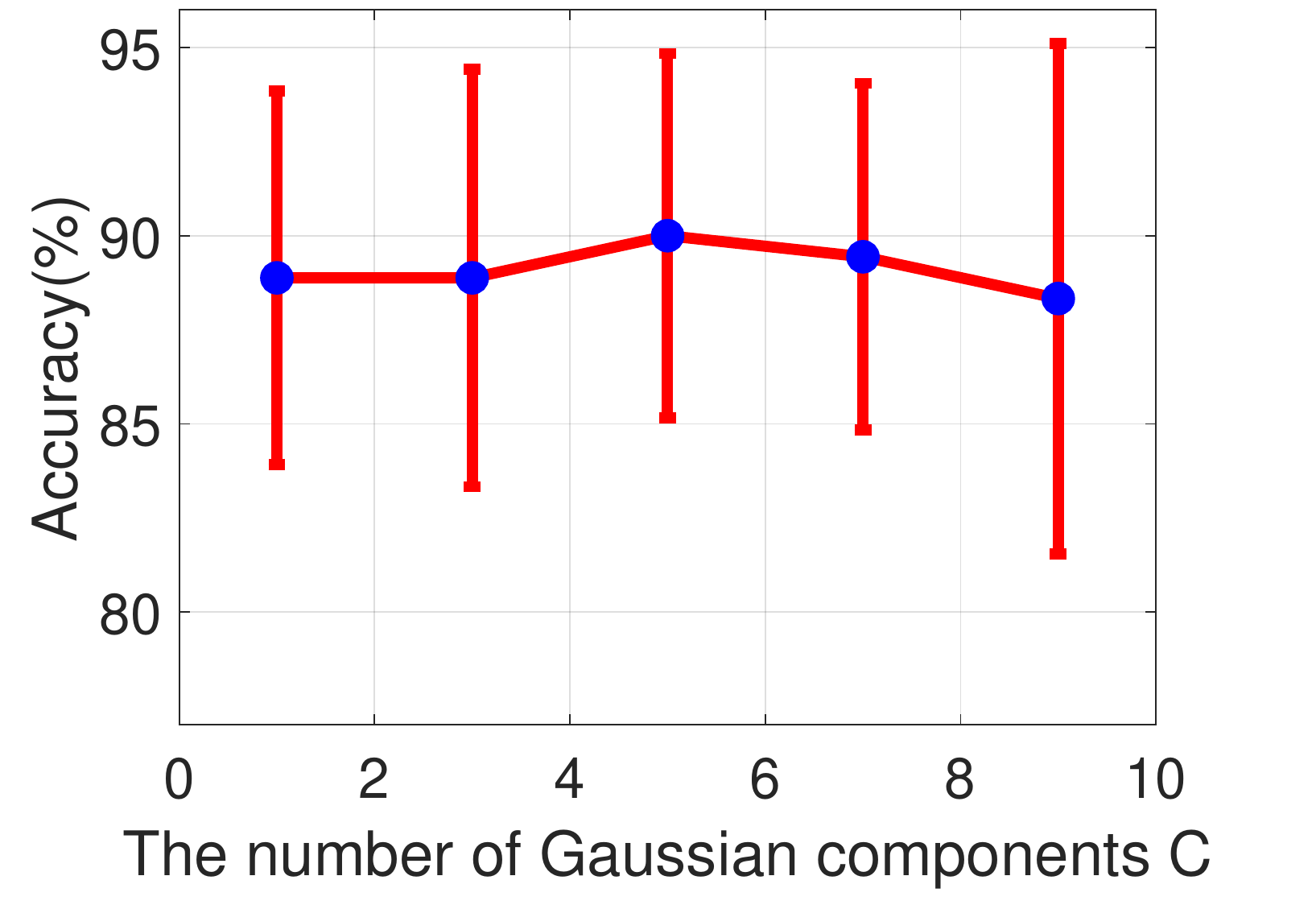}}
	    \subfigure[\Chunyan{With variable $T$ and fixed $C=9$}]{
			\label{fig:ct_t}
			\includegraphics[width=0.23\textwidth]{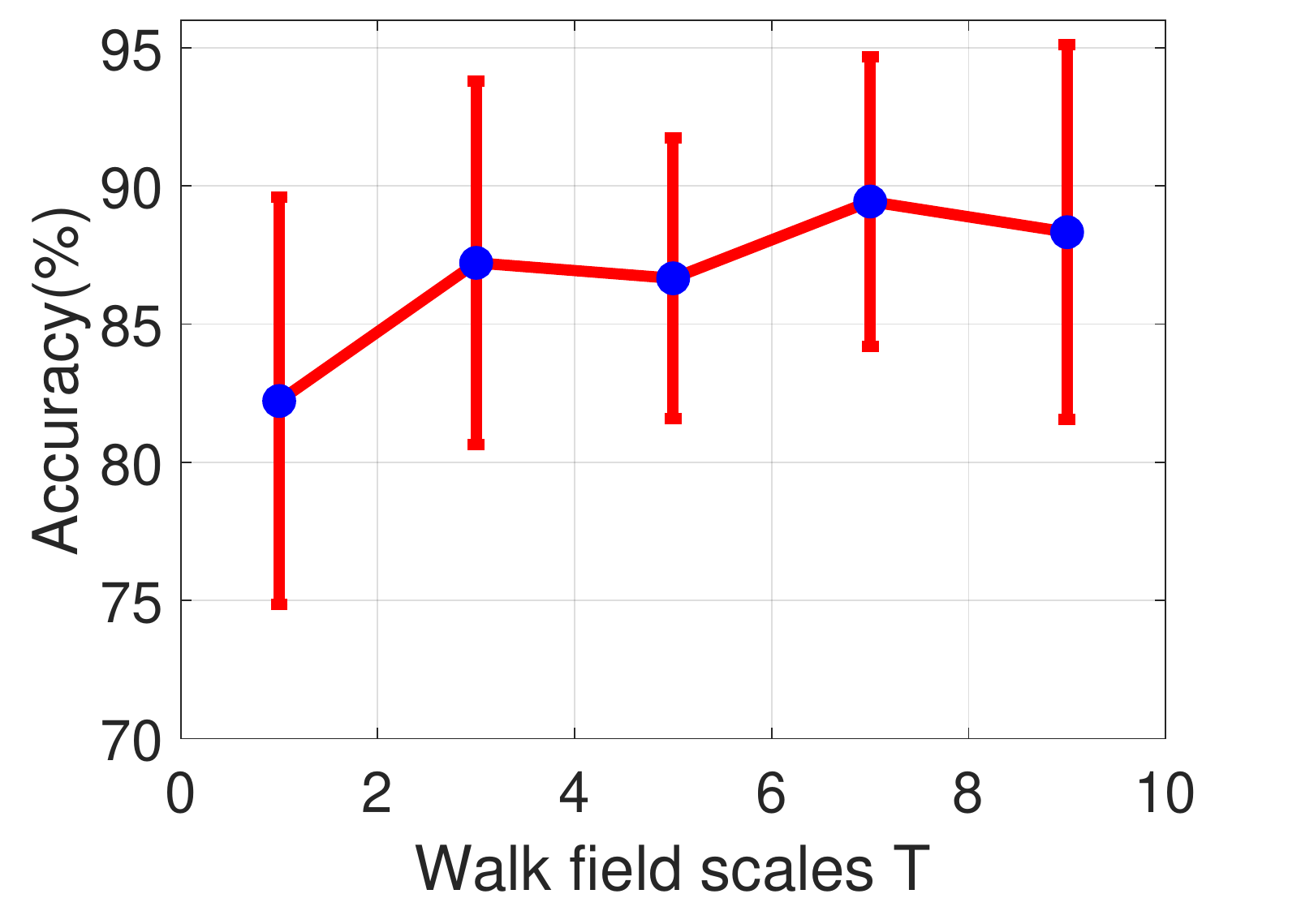}}
	\end{center}
	\vskip -0.1in
	\caption{\JJ{The graph classification performance of our proposed WSC with different values of $C$ and $T$ on MUTAG dataset. 
	$C$ and $T$ denote the number of Gaussian components and the scale of walk fields in the graph convolution operator, respectively. 
	 }}
	\label{fig:parameter1}
	\vskip -0.1in
\end{figure}

\begin{figure*}[!t]
	\begin{center}
		\centering
        \subfigure[\Chunyan{on PROTEINS dataset}]{
			\label{fig:rf_proteins}
			\includegraphics[width=0.3\textwidth]{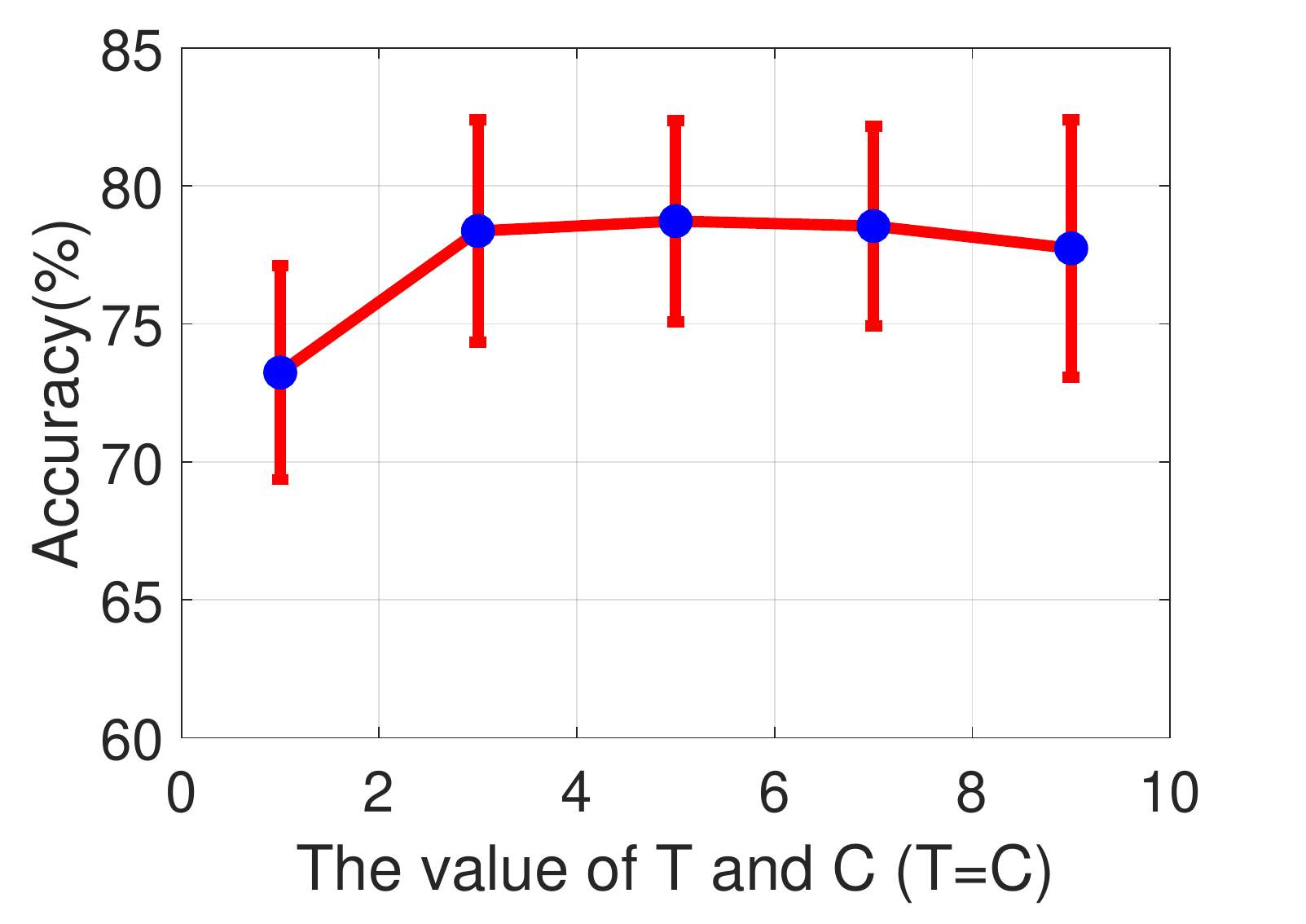}}
		\subfigure[\Chunyan{on PTC dataset}]{
			\label{fig:rf_ptc}
			\includegraphics[width=0.3\textwidth]{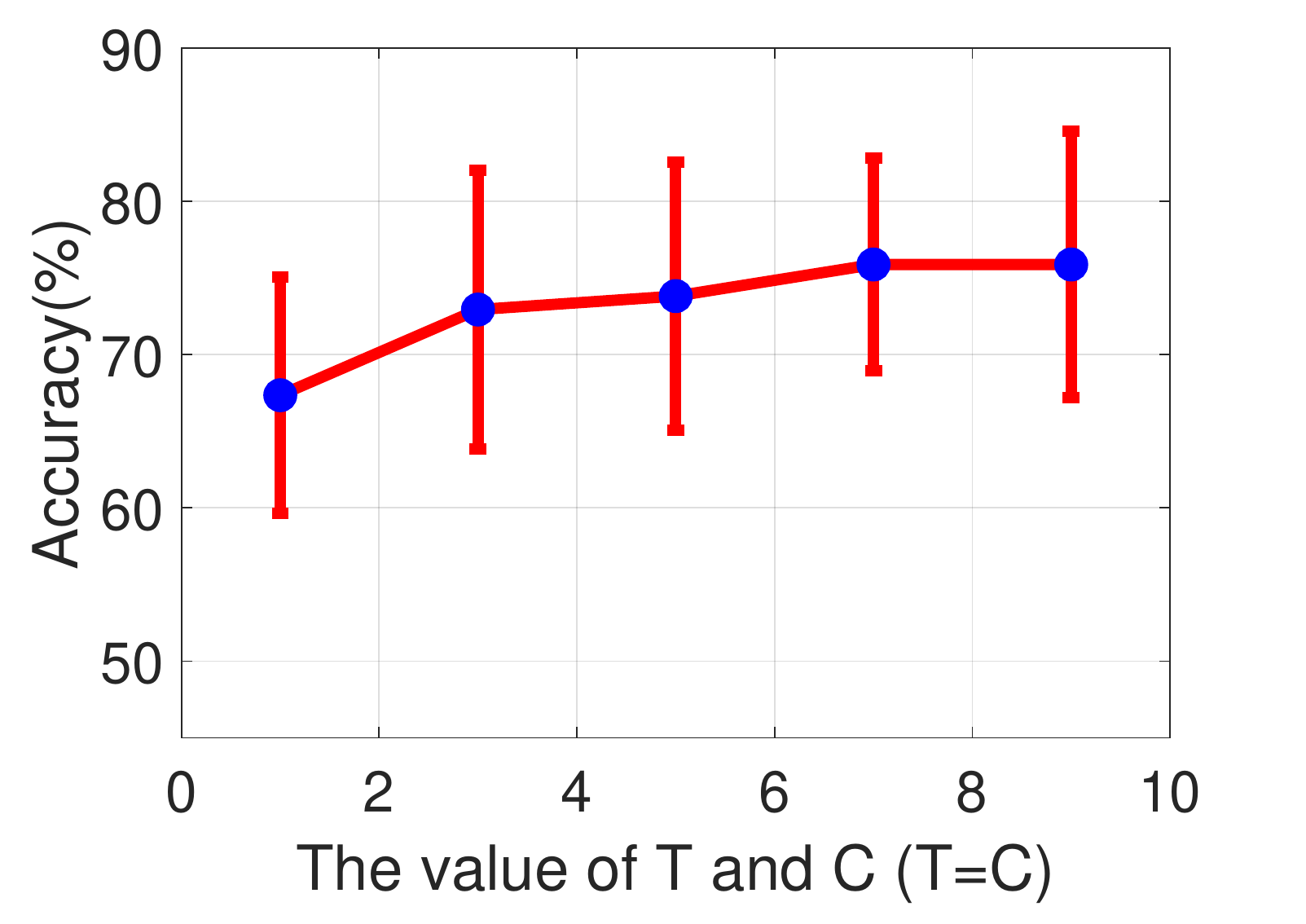}}
        \subfigure[\Chunyan{on MUTAG dataset}]{
			\label{fig:rf_mutag}
			\includegraphics[width=0.3\textwidth]{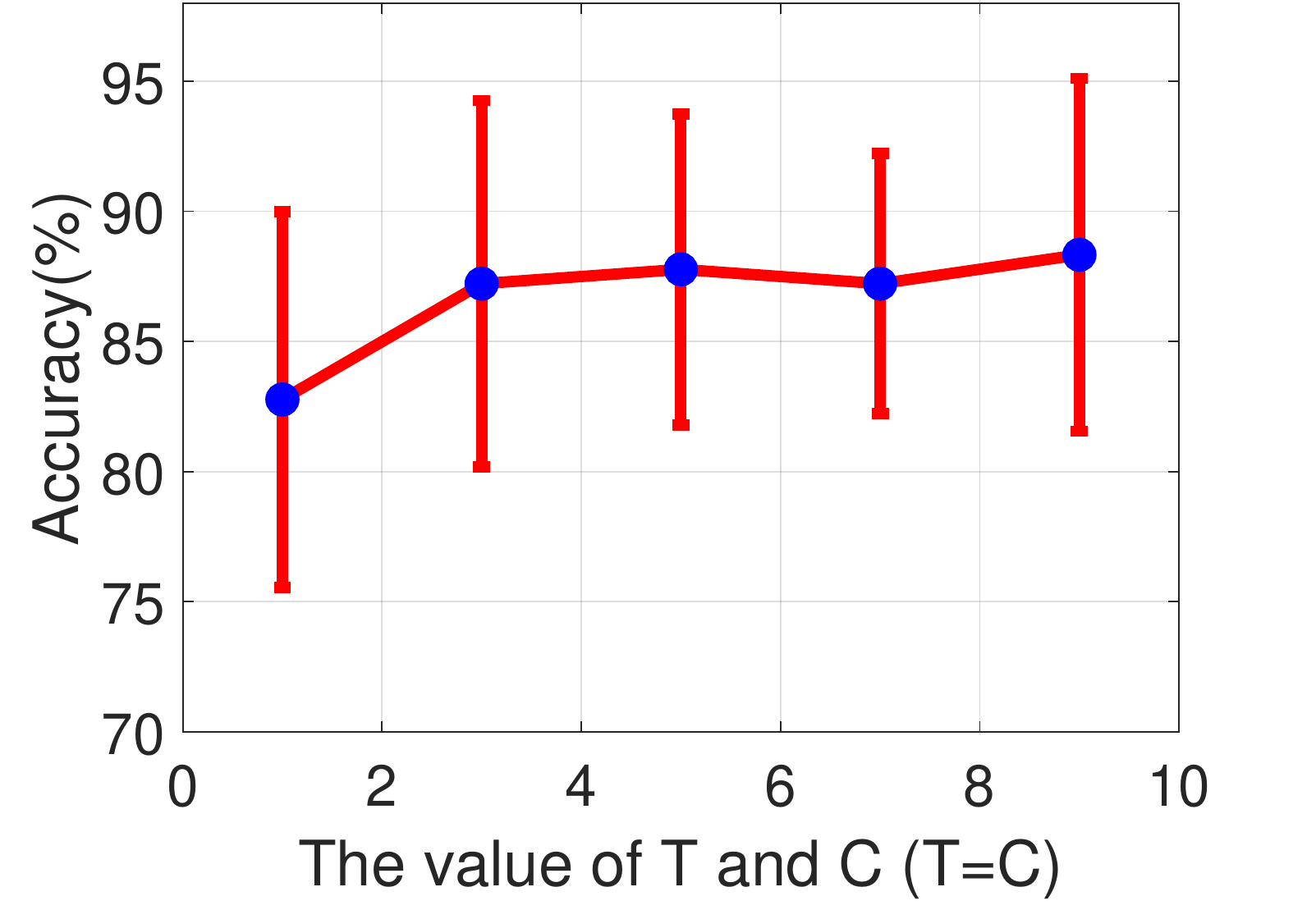}}
	\end{center}
	\vskip -0.1in
	\caption{\JJ{The graph classification performance with jointly varying walk scales and Gaussian components, \ie, $\{(T,C)|T=C\}$, on PROTEINS, PTC and MUTAG datasets.} 
	}
	\label{fig:parameter2}
	\vskip -0.1in
\end{figure*}

\begin{figure*}[!t]
	\begin{center}
		\centering
        \subfigure[\Chunyan{on PROTEINS dataset} ]{
			\label{fig:N_ptc}
			\includegraphics[width=0.3\textwidth]{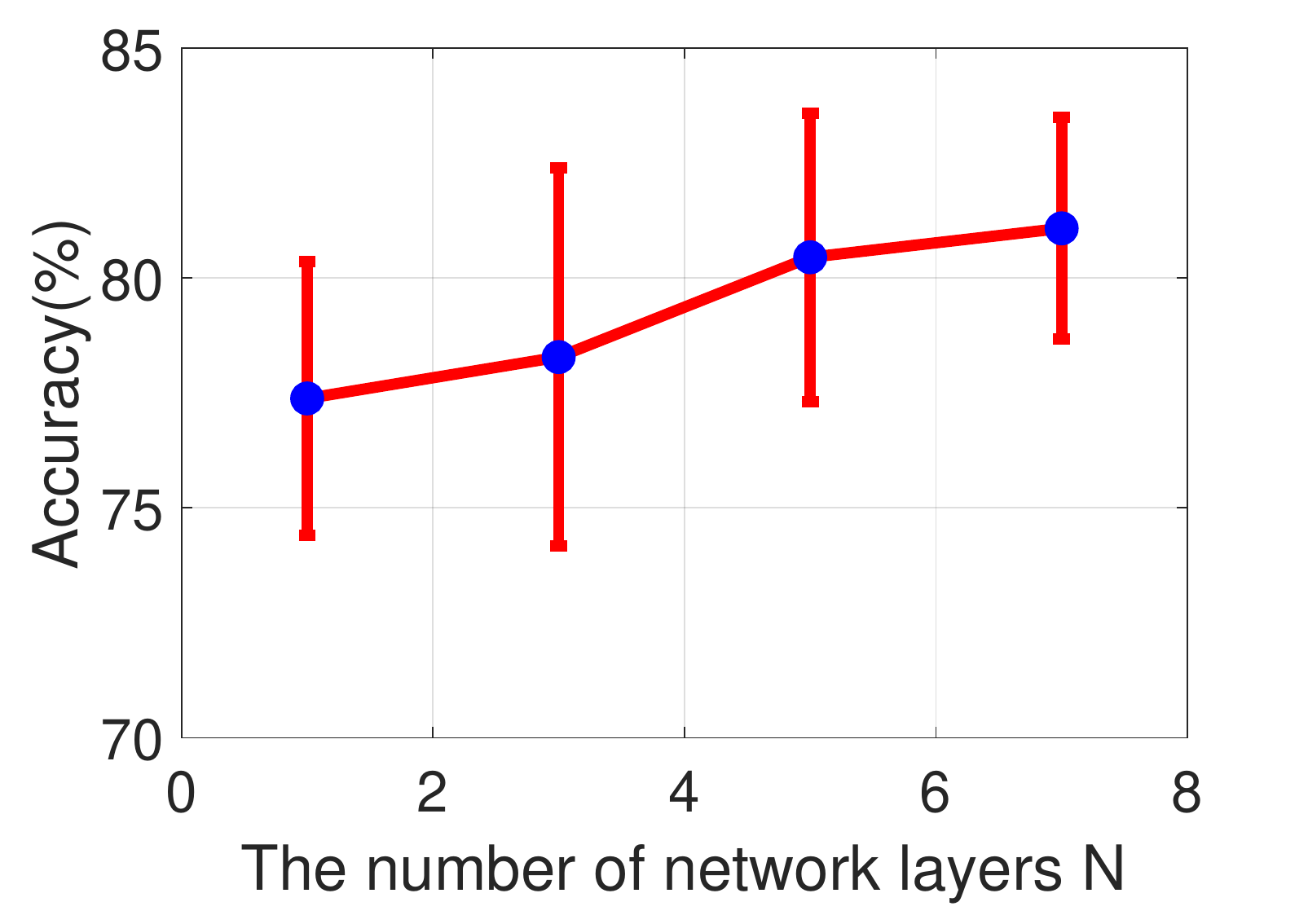}}
        \subfigure[\Chunyan{on PTC dataset} ]{
			\label{fig:N_proteins}
			\includegraphics[width=0.3\textwidth]{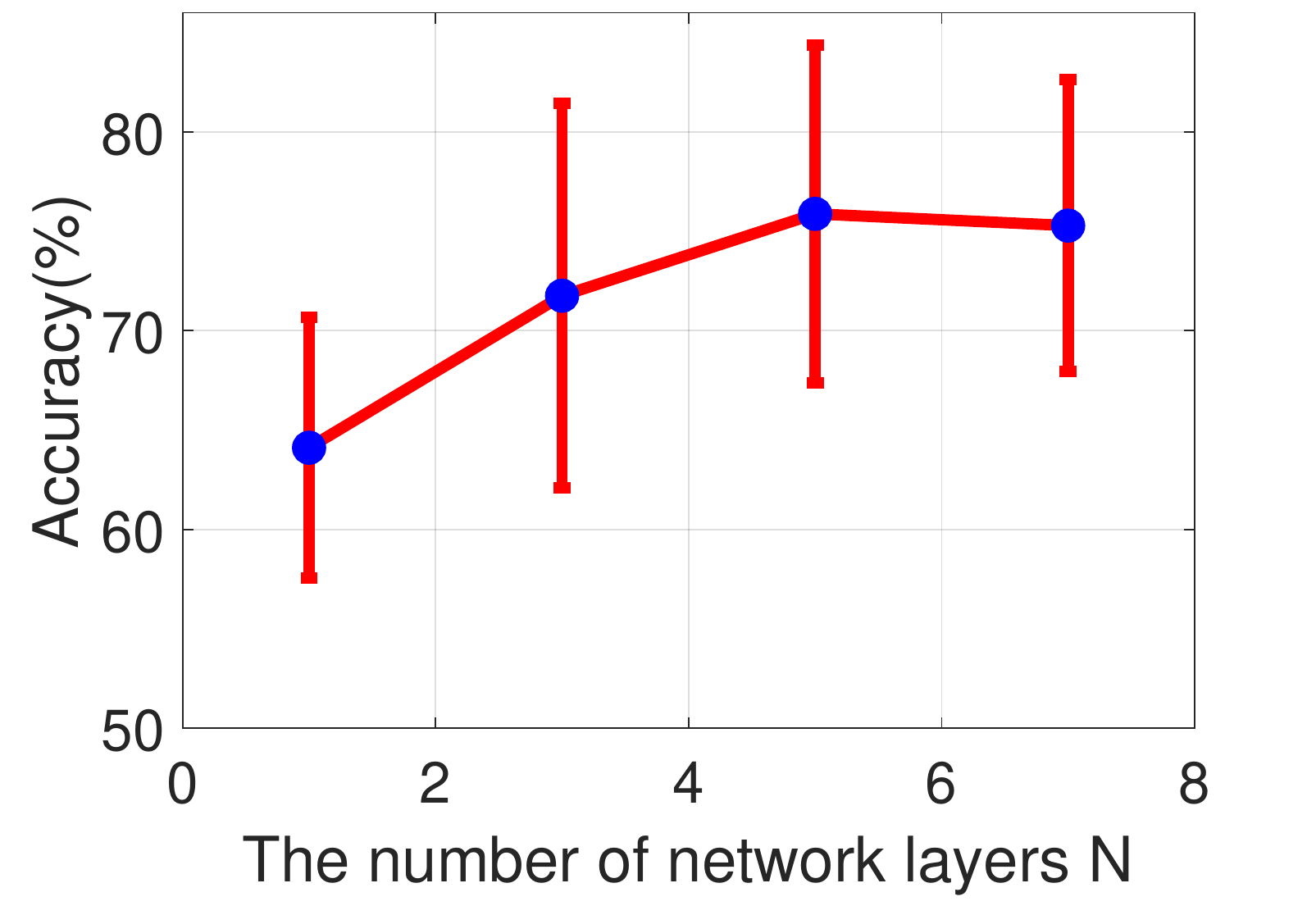}}
        \subfigure[\Chunyan{on MUTAG dataset}]{
			\label{fig:N_mutag}
			\includegraphics[width=0.3\textwidth]{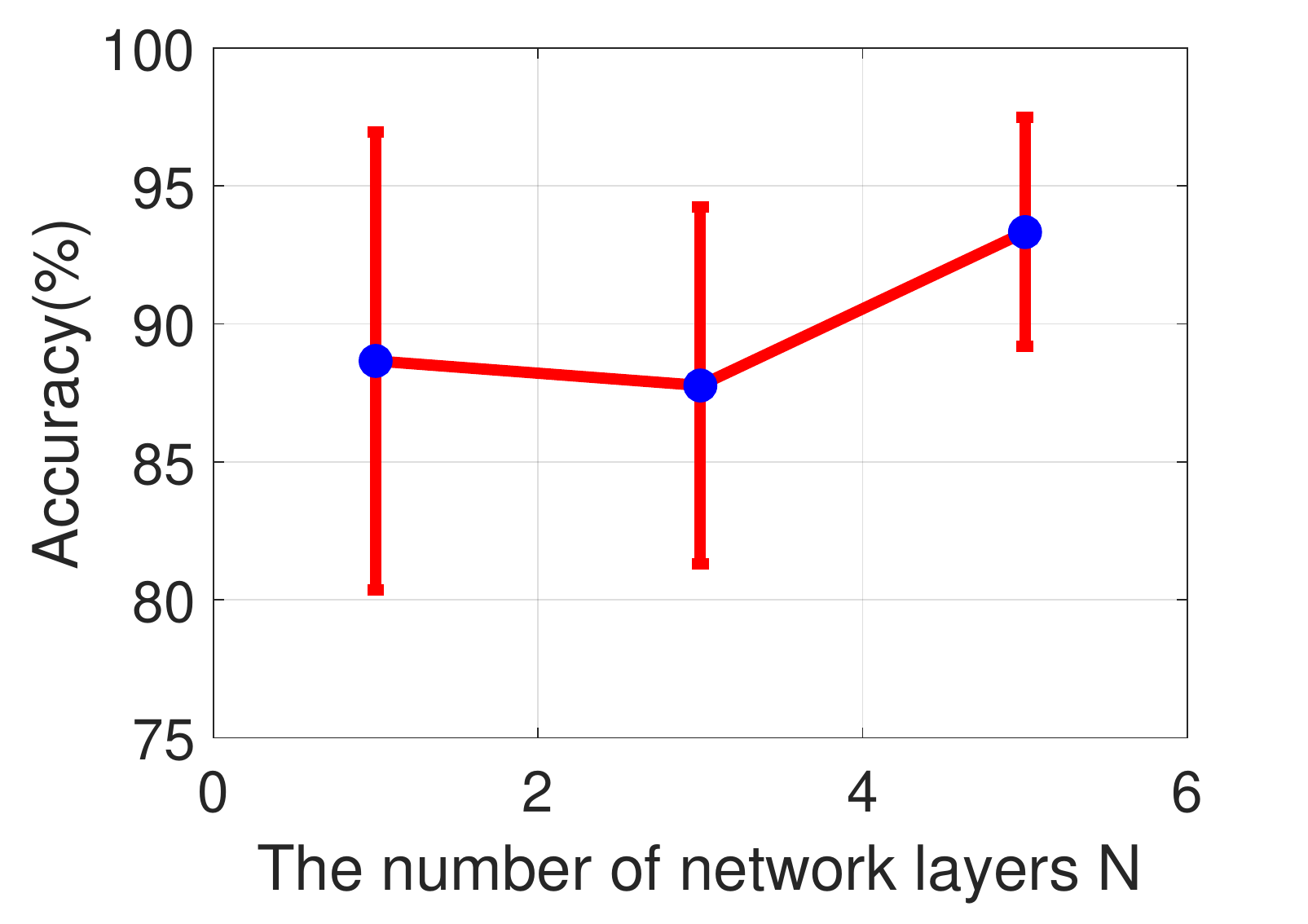}}
	\end{center}
	\vskip -0.1in
	\caption{\JJ{The graph classification performance with different numbers of network layers $N$ on PROTEINS, PTC and MUTAG datasets.}}
	\label{fig:parameter3}
	\vskip -0.1in
\end{figure*}

\begin{figure*}[!h]
	\begin{center}
		\centering
		\subfigure[\Chunyan{on PROTEINS}]{
			\label{fig:K_proteins}
			\includegraphics[width=0.3\textwidth]{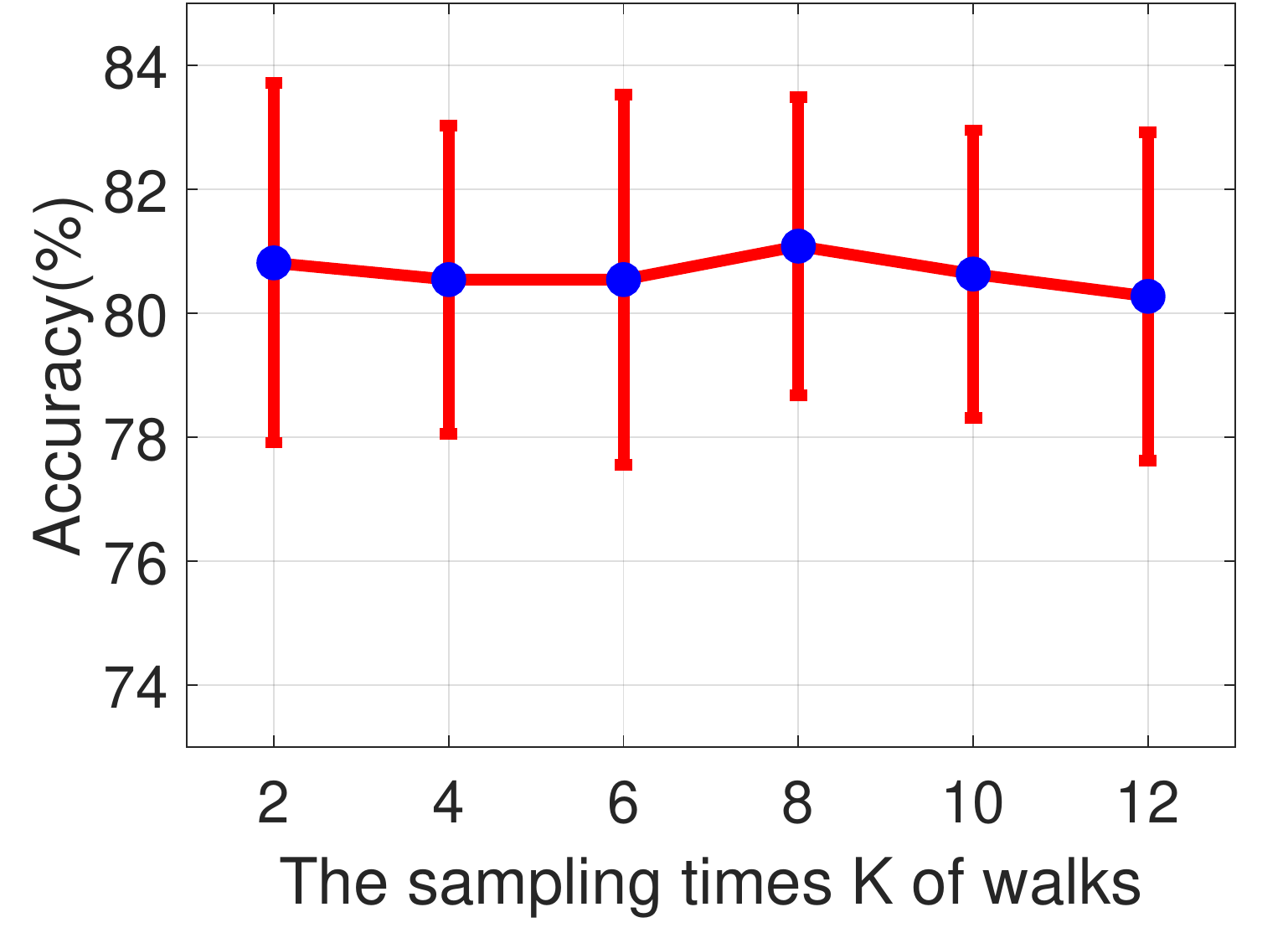}}
		\subfigure[\Chunyan{on PTC dataset}]{
			\label{fig:K_ptc}
			\includegraphics[width=0.3\textwidth]{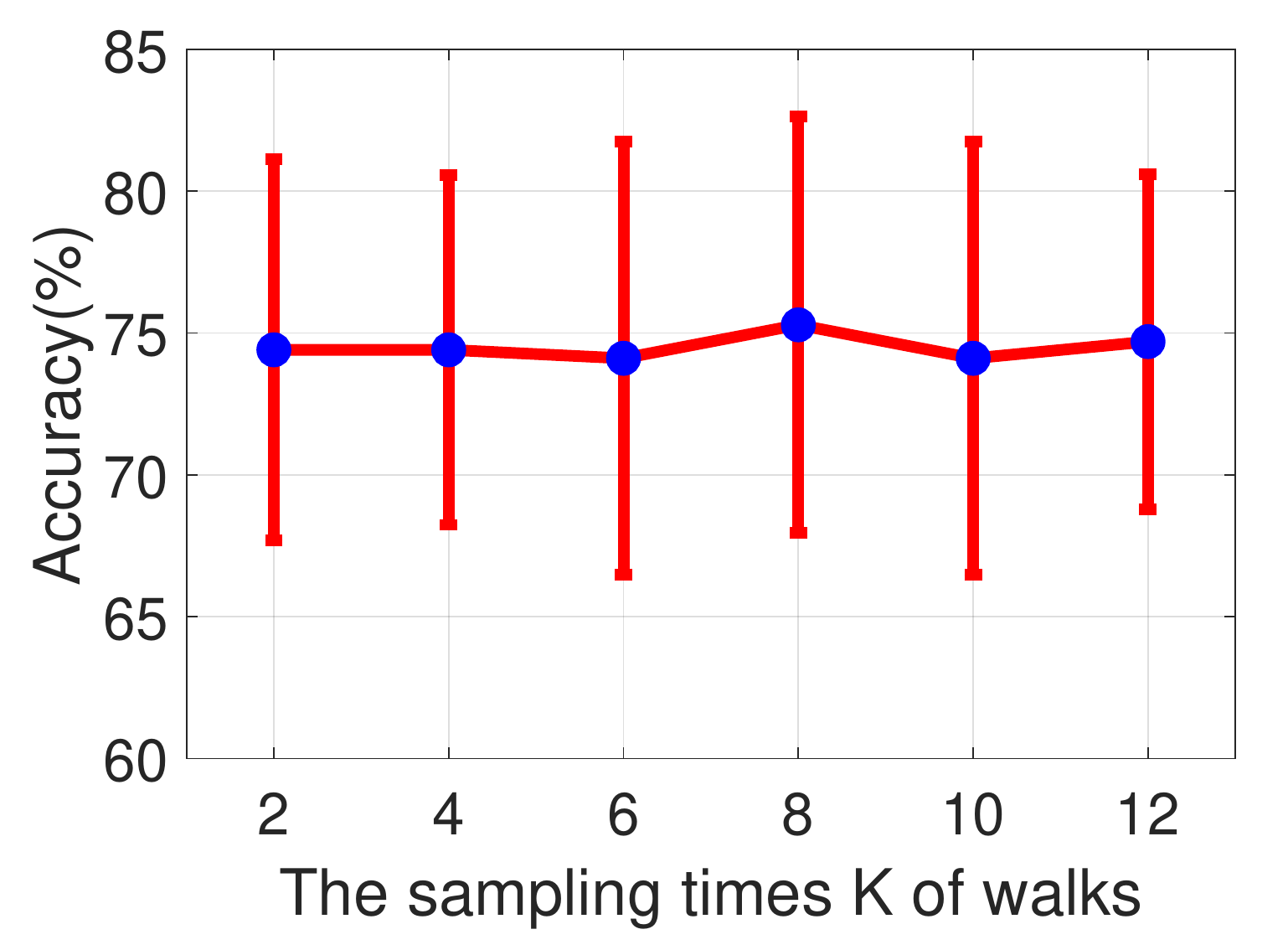}}
		\subfigure[\Chunyan{on MUTAG dataset}]{
			\label{fig:K_mutag}
			\includegraphics[width=0.3\textwidth]{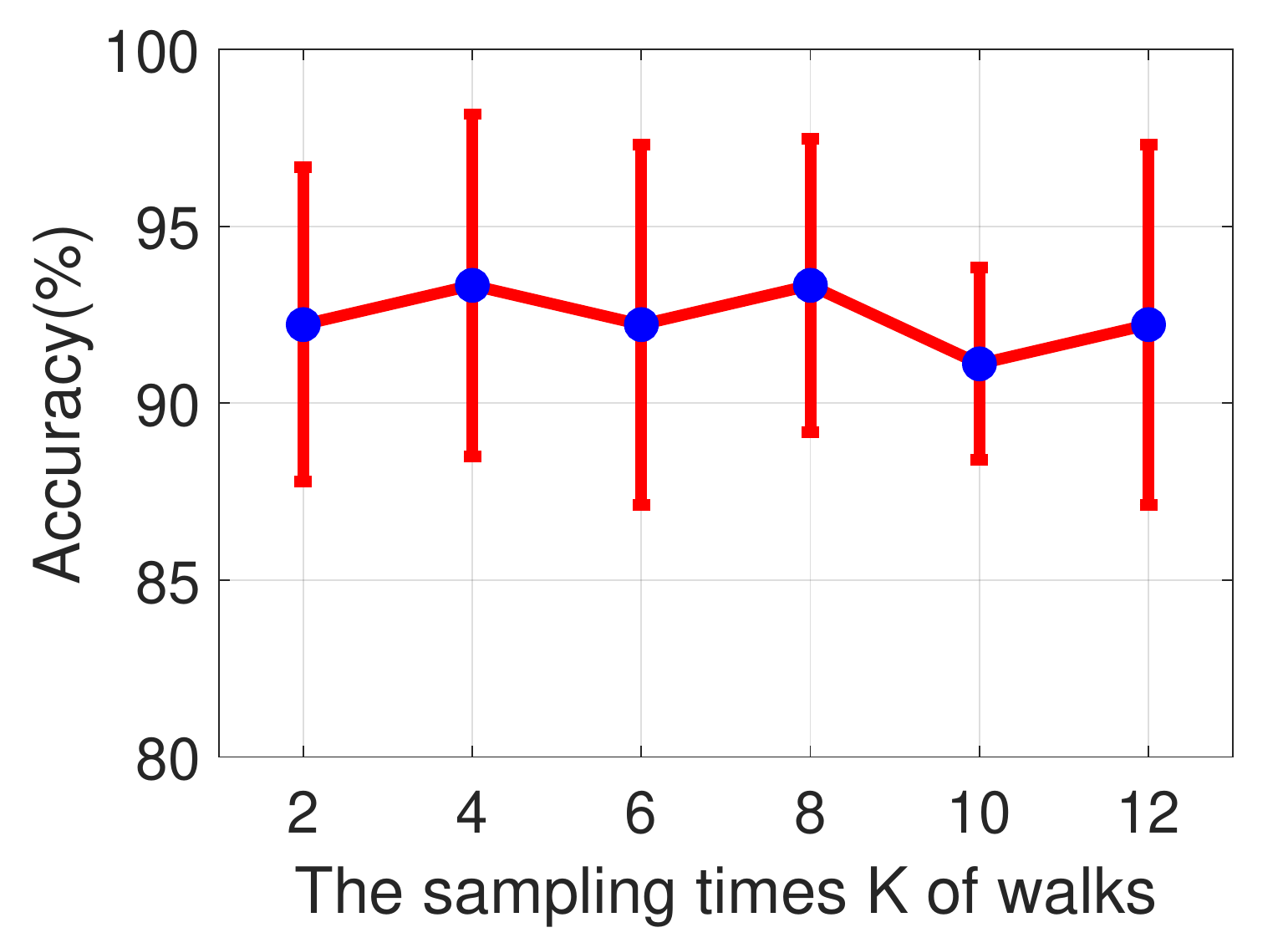}}
	\end{center}
	\vskip -0.1in
	\caption{\JJ{The graph classification performance with different sampling times $K$ of walks on PROTEINS, PTC and MUTAG datasets.}}
	\label{fig:sampling}
	\vskip -0.2in
\end{figure*}

We firstly evaluate the performance of our WSC with different parameters of $C$ and $T$ on MUTAG dataset, where $C$ and $T$ denotes the number of Gaussian components and the scale of walk fields in the graph convolution operator, respectively. Suppose the scale of walk fields $T$ is fixed to 9, then we vary Gaussian components $C$ from 1 to 9 with interval 2. As shown in Fig.~\ref{fig:ct_c}, we can observe that the classification accuracies are first improved and then declined. The reason is that graph convolution with less Gaussian components may not be enough to encode local variations of walk fields during the convolution filtering, but more Gaussian components might result into the overfitting. The overfitting comes from two folds:
i) the total number of Gaussian models is $T\times C$ for each convolution layer when each walk field scale has $C$ Gaussian models;
ii) the filtering parameter space is giant in Eq.~(\ref{eqn:filtering}) especially for small-sample datasets.
Oppositely, when fixing $C$ to 9, the increase of $T$ can generally promote the performance as shown in Fig.~\ref{fig:ct_t}, as more global information can be encoded during convolutional filtering. However, a larger $T$ will lead to  heavier computation burden. Therefore, a balance between Gaussian components and the computation complexity of network parameters can get a better recognition accuracy in practice.

In view of the fact that larger-scale random walks cover bigger local subgraphs with more complex variations, the number of Gaussian components should be increased to encoded those variations well. Hence,
we let $C$ be proportional to $T$, simply set $C=T$. The experimental verifications on their joint changes are shown in Fig.~\ref{fig:rf_proteins}, \ref{fig:rf_ptc} and \ref{fig:rf_mutag}, where only one convolutional layer (e.g., C(64)-P(0.0)-FC(256)) is employed to demonstrate their influences frankly. Under this configuration, we can observe that the performances are basically ascendant with the increase of their values. However, larger values on $C, T$ would burden the computation process, or sometimes result into overfitting according to the above analysis. Practically the setting of $C, T =3$ is a good tradeoff between computational efficiency and decent results.

We conduct the experiments on the number of stacked convolutional layers. \Chunyan{In order to test how the number of network layers influence the performance, we compare  four different networks, where detailed configurations are FC(256), C(64)-P(0.0)-FC(256), C(64)-P(0.25)-C(128)-P(0.0)-FC(256) and C(64)-P(0.25)-C(128)-P(0.25)-C(256)-P(0.0)-FC(256), respectively.}
The detailed comparisons are shown in Fig.~\ref{fig:N_proteins}, \ref{fig:N_ptc} and \ref{fig:N_mutag}. For MUTAG dataset, we do not employ three convolution layers as shown in Fig.~\ref{fig:N_mutag}. Because graph data have less vertices on MUTAG datasets, which is not enough to perform three coarsening operators with 0.25 ratio. More details can be found in Section~\ref{sec:datasets} and Section~\ref{sec:details}.
For example, the classification results on PROTEINS dataset are 77.38\%, 78.28\%, 80.45\% and 81.08\%, respectively.
We also observe that the graph network with three-layer convolution filtering can achieve a comparable performance on all graph datasets.
Of course, for graph data with millions of vertices, the number of layers should be further increased to model the complex data like conventional deep neural networks.

\begin{table}[!t]
	\centering
	\caption{\Chunyan{Comparisons of our WSC performance with different graph attributes.}}
	\setlength{\tabcolsep}{4pt}
	\label{table:nodegree}
	\scalebox{1}{
	\begin{tabular}{l c c c c}
		\hline
		Dataset  & WSC w/ label   & WSC w/ degree    &WSC w/ label \& degree \\
		\hline
		MUTAG          & 90.00 $\pm$ 5.98      & \textbf{95.55 $\pm$ 4.15}  & 93.33 $\pm$ 4.15 \\
		PTC            & 73.52 $\pm$ 6.83      & 74.11 $\pm$ 5.22     & \textbf{75.29 $\pm$ 7.34} \\
		NCI1           & 79.56 $\pm$ 0.75      & 78.41 $\pm$ 1.19     & \textbf{81.70 $\pm$ 0.98}	 \\		
		ENZYMES        & 56.16 $\pm$ 7.07      & 51.50 $\pm$ 4.62     & \textbf{56.16 $\pm$ 5.91} \\
		PROTEINS       & 79.54 $\pm$ 3.01      & 79.36 $\pm$ 2.77     & \textbf{81.08 $\pm$ 2.41} \\
		\bottomrule
	\end{tabular}
	}
	\vskip -0.1in
\end{table}

\Chunyan{
We further evaluate graph classification accuracies of our proposed WSC with different sampling times $K$ of walks on PROTEINS, PTC and MUTAG datasets, and the detailed results can be shown in Fig.~\ref{fig:sampling}. 
The experimental results indicate that the performance of WSC can be slightly affected with different sampling times $K$ of walks, and the WSC can gain a comparable performance on PROTEINS, PTC and MUTAG graph datasets when $K=8$. 
Therefore, in all our experiments, the default sampling times $K$ of walks can be set to 8 for achieving a better graph classification performance of WSC. 
}
\Chunyan{
To compare the WSC performance with different graph attributes, we use the same network settings of our WSC, but with different graph attributes (including label, degree and both information), named ``WSC w/ label", ``WSC w/ degree" and ``WSC w/ label \& degree", respectively. 
As can be seen in Table~\ref{table:nodegree}, the ``WSC w/ label \& degree" can perform better than WSC only with degree or label information (i.e., ``WSC w/ label" and ``WSC w/ degree") on all bioinformatics datasets except MUTAG. 
It demonstrates that our WSC performs very well with both label and degree information, but its performance would be slightly decreased with only vertex degree or label information on most graph datasets, including PTC, NCI1, ENZYMES and PROTEINS bioinformatics datasets.}



\subsection{Computation Complexity Analysis}

Here we do an analysis of the computation complexity of our proposed WSC network, which mainly depends on the walk fields construction and Gaussian mixture encoding.
The amount of computation is approximately linear to the graph vertex numbers $m$ on sparse graphs, as the random walk scale $T$ and random walk times $K$ are far less than $m$.
The computation complexity of Gaussian mixture encoding (including attribute mapping) is about $O(KTCd + KTCd^2)$ for each convolution operator, where $d$ is the dimension of node attributes.
The computation cost of convolution operator is lower than those spectral methods, which is about $O(m^3)$ for eigenvalue decomposition on graph Laplacian matrix. This is even comparable to the approximation methods~\cite{defferrard2016convolutional}, which is about $O(m^2)$ due to $m > d$ in most cases. 
\begin{figure}[!t]
	\begin{center}
		\centering
		\subfigure[The running times of training process with 400 epochs]{
			\includegraphics[width=0.23\textwidth]{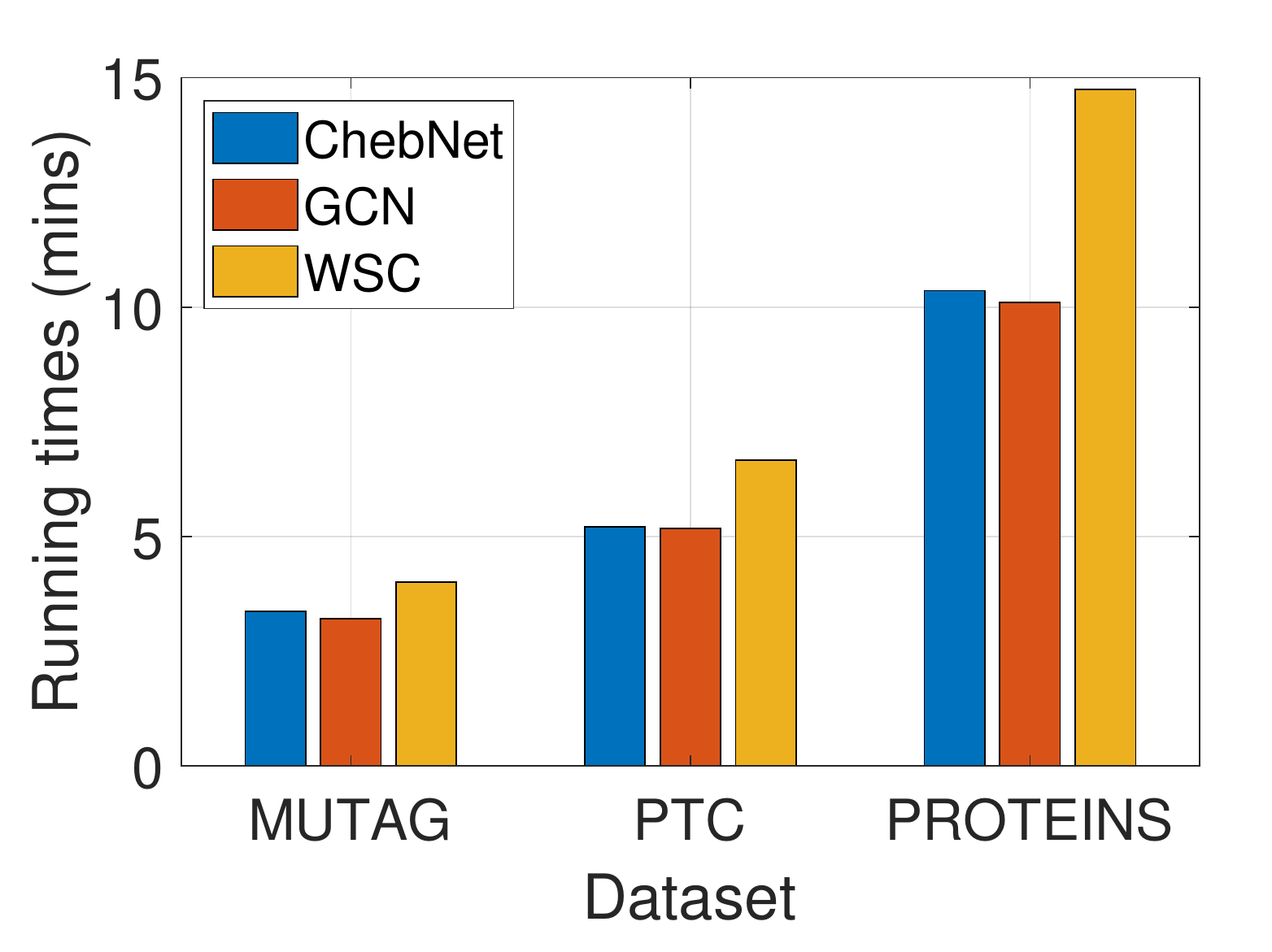}}
		\subfigure[The running times of testing process]{
			\includegraphics[width=0.23\textwidth]{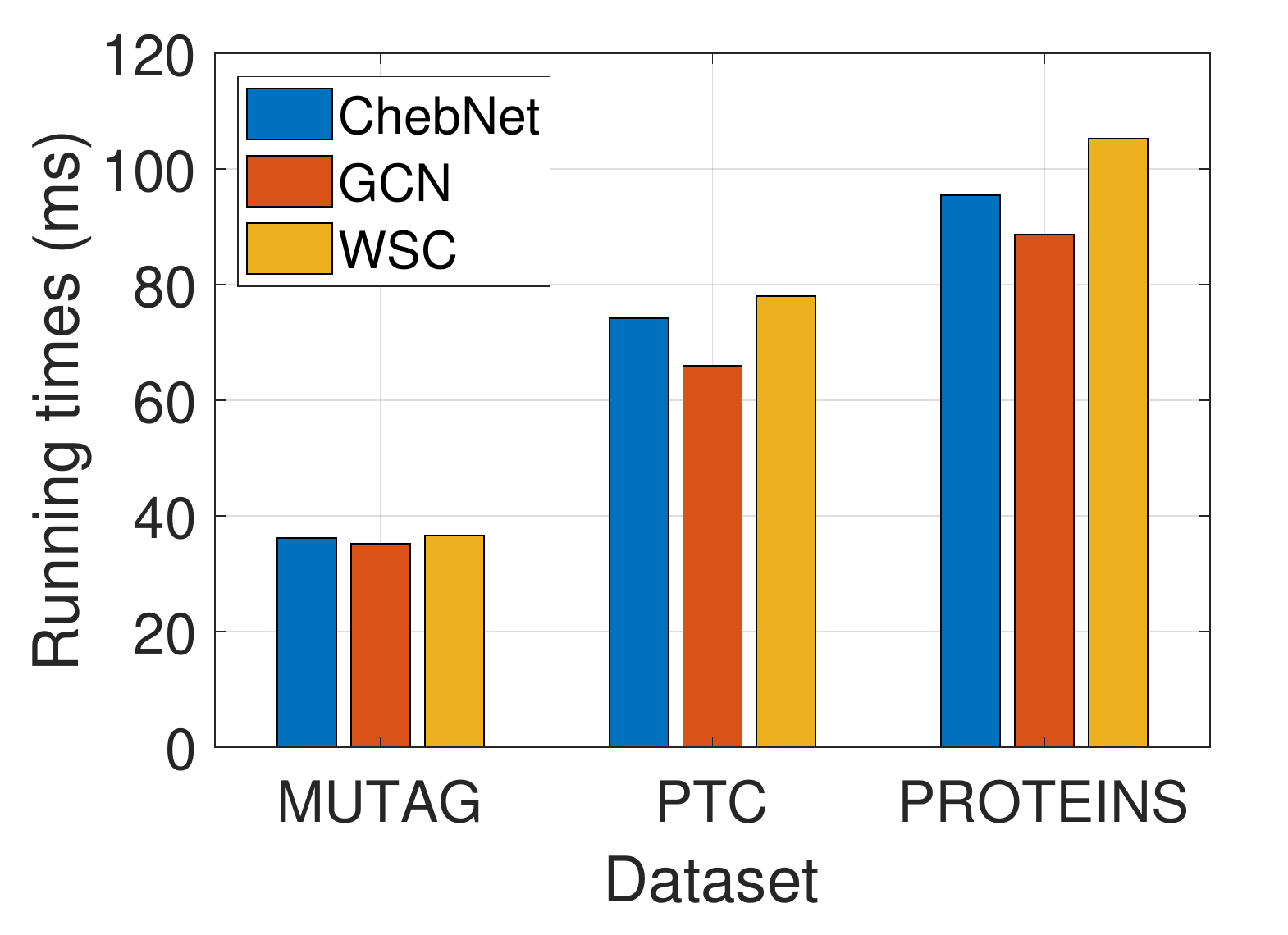}}
	\end{center}
	\vskip -0.1in
	\caption{\JJ{The running times of ChebNet, GCN and WSC on MUTAG, PTC and PROTEINS datasets.}}
	\label{fig:runtime}
	\vskip -0.1in
\end{figure}
\JJ{Further, we compare the proposed model with these baseline methods (including ChebNet~\cite{defferrard2016convolutional}, GCN~\cite{kipf2016semi}) in the running time of training and testing processes, where the same network settings and experimental environment (a single GPU of RTX 2080ti) but different graph convolution modules are adopted with 400 epochs on MUTAG, PTC and PROTEINS datasets. As can be seen in Fig.~\ref{fig:runtime}, our method spends a little more time than these baseline methods. The extra time mainly comes from the calculation of projections on Gaussian models, but it accounts for a small ratio in the entire process. The experiment comparison shows that our method is also efficient in practice. Furthermore, the experiments on running time are also in accord with the above computation complexity analysis.}

\section{Conclusion}

In this paper, we proposed a walk-steered convolution network to better represent these irregular graphs, and applied to the graph classification task.
Specifically, we introduced a novel graph convolution method, which built multi-scale walk fields to define local receptive fields, and then employed Gaussian mixture models to encode local variations of each receptive field on graph-structured data.
Further, we employed a \Jiang{graph clustering algorithm} to coarsen graphs for simultaneously abstracting hierarchical structures of graphs and reducing the computational cost.
Different from previous graph convolution methods, our proposed WSC can be more flexibility to graphs and well encode variations of local subgraph regions.
Experimental results on several public graph datasets validated the effectiveness and superiority of our WSC network.
\Chunyan{In the future, we would like to extend our methods for large-scale graph data and explore more applications to graph understanding, e.g., social network, traffic control and chemical reaction analysis.} 



%

%


\ifCLASSOPTIONcaptionsoff
  \newpage
\fi



%

%
%

\begin{small}
  \bibliographystyle{IEEETran}
  \bibliography{myref}
\end{small}
\vskip -0.6in

\end{document}